\documentstyle[12pt,fleqn]{article}
\catcode`\@=11
\@addtoreset{equation}{section}
\def\ksection{\arabic{section}}
\def\theequation{\ksection.\arabic{equation}}
\def\t{{t}}
\def\ti{{\t_{\rm i}}}

\def\DR{{\Delta R(\ti)}}

\def\Rmx{{R_B}}

\def\sG{{\sigma_{\Gamma}}}
\def\RW{{R_{\rm w}}}
\def\pW{{p_{\rm w}}}
\def\rhoW{{\rho_{\rm w}}}
\def\UW{{U_{\rm w}}}
\def\PhiW{{\Phi_{\rm w}}}
\def\GW{{\Gamma_{\rm w}}}
\def\rw{{\Delta R_{\rm w}}}
\def\Hm{{H^{-1}_{\rm out}(\ti)}}
\def\rhoo{{\rho_{\rm out}(\ti)}}
\def\rhoi{{\rho_{\rm in}(\ti)}}
\def\w{{\omega}}
\def\chit{{\widetilde{\chi}}}
\def\pp{{\widetilde{p}}}

\def\XW{{X_{\rm w}(\ti)}}

\def\rhob{{\overline{\rho}}}

\def\Phib{{\overline{\Phi}_{\rm in}}}
\def\x{{(R-\RW)/\rw}}

\def\sig{{\sigma}}
\def\ep{{\epsilon}}

\def\alp{{\alpha}}
\def\alpp{{\tilde{\alpha}}}
\def\bet{{\beta}}

\def\Lam{{\Lambda}}
\def\f{{\overline{f}}}

\def\rs{{\dot{r}_{\rm sh}}}

\def\dr{{\Delta R(\ti)}}
\def\b{{\rm b}}

\def\dn{{\dot{n}}}
\def\dR{{\dot{R}}}
\def\dM{{\dot{M}}}
\def\dU{{\dot{U}}}
\def\drho{{\dot{\rho}}}
\def\dep{{\dot{\ep}}}
\def\der{{\nabla}}

\def\pa{{\partial}}
\def\be{\begin{equation}}
\def\ee{\end{equation}}
\def\ba{\begin{eqnarray}}
\def\ea{\end{eqnarray}}
\def\la{\mathrel{\mathpalette\fun <}}
\def\ga{\mathrel{\mathpalette\fun >}}
\def\fun#1#2{\lower3.6pt\vbox{\baselineskip0pt\lineskip.9pt
        \ialign{$\mathsurround=0pt#1\hfill##\hfil$\crcr#2\crcr\sim\crcr}}}

\def\re#1{{[\ref{#1}]}}
\def\eqr#1{{Eq.\ (\ref{#1})}}
\begin{document}
\begin{titlepage}
\null\vspace{-62pt}
\renewcommand{\thefootnote}{\alph{footnote}}
\begin{flushright}CfPA 94-th-23\\
September, 1994\\
{\it Accepted by Phys.~Rev.~D August 29, 1994}
\end{flushright}
\vspace{0.5in}
\vspace{0.1in}
\begin{center}
{\Large \bf Collapse of a Superhorizon-sized Void\\ in the
Early Universe}\\
\vspace{1.0cm}
Sharon L.\ Vadas\footnote{Electronic mail: {\tt
vasha@physics.berkeley.edu}}\\
{\em Center for Particle Astrophysics\\
University of California, Berkeley, CA~~94720}\\
\end{center}

\baselineskip=24pt

\vspace{1.5cm}

\begin{quotation}

In this paper, we study the collapse of a superhorizon-sized void in
the early, radiation-dominated
universe using an improved
general relativistic code.
We find that in general a relativistic or nonrelativistic void collapses
via a shock at the speed of light.
This is
true even if the outward velocity of the void
wall is enormous.  As the wall thickness
decreases, the shock strength increases and the collapse time decreases
up to a limit set by the shock tube solution.
In addition, as the outward velocity of the wall increases, the collapse time
increases somewhat.
When the collapse occurs in much less than a Hubble time outside the void,
gravitational forces contribute negligibly to the solution at the
collapse time; non-gravitational forces (caused by pressure
and velocity gradients)
shape the solution almost entirely.  This is true even if the wall velocity
is large enough that fluid in the peak shocks outward substantially
during this time.
Gravitational forces are expected to be dominant only {\it after}
a void has collapsed.

\vspace*{12pt}

PACS number(s): 98.80.Cq, 04.25.Dm, 47.55.Dz

\end{quotation}
\end{titlepage}
\newpage
\baselineskip=14pt

\centerline{\bf{I.  Introduction}}
\setcounter{section}{1}
\setcounter{equation}{0}
\renewcommand{\thefootnote}{\alph{footnote}}

Because the total energy density is of order the critical energy
density today,
if the early universe is described entirely by the big bang model,
a severe fine-tuning problem exists.
This problem is solved if inflation occurred in
the early universe.
First-order
inflation is one of the most interesting classes of inflationary
models\re{LASTEIN}.  Here, inflation ends in a particular
region when a true-vacuum bubble is nucleated.
Inflation ends everywhere when the universe is filled
with true vacuum bubbles.
At this point, scalar
field dynamics occurs which creates relativistic, fundamental particles.
These particles thermalize much of the universe within a few
Hubble times.
Due to causality however, bubbles
nucleated early on during inflation cannot be thermalized as quickly.
Thus, since the inside of these bubbles are
virtually empty, superhorizon-sized
voids are created
at the end of reheating.\footnote{Following past
convention, we loosely equate the Hubble radius
with the horizon in the phrase ``superhorizon-sized''.
Horizon in this context
is not to be confused with the particle horizon.}
These voids are larger than the outside Hubble radius, $c\Hm$, (i.e. the Hubble
radius outside the void), and can have enormous
radii $\RW\sim 10^{27}c\Hm$.
The beginning of the radiation-dominated epoch then, is characterized
by an approximately flat Friedmann-Robertson Walker (FRW)
homogeneous and isotropic universe, punctuated infrequently
by nearly
empty, superhorizon-sized voids.

The thermalization of the very large voids has been thought to take an enormous
amount of time and therefore be in conflict with the tiny temperature
fluctuations
measured in the microwave background\re{BigBubble}.  In a previous
paper\re{Vasha}
however (hereafter referred to as SV), it was
found that these former estimates could be incorrect because
the first-crossing time
(the time taken for a photon in the void wall to reach the origin
and thus make ``contact'' with a photon from the other side) was
calculated incorrectly.
In \re{BigBubble}, it was found that the first-crossing time, $\Delta t_c$,
(i.e. the minimum thermalization time) is
$\Delta t_c/\Hm\simeq (c^{-1}\RW/\Hm)^2$.
However,
it was found in SV that the first-crossing time
is actually much smaller than this:
\ba
\label{eq:DeltHGR}
{\Delta t_c}/{H^{-1}_{\rm out}(\ti)}=\Phib^{-1}({c^{-1}\RW}/{\Hm})
\left[1+.5\Phib^{-1}({c^{-1}\RW}/{\Hm})\right],
\ea
where the void's relative size is $c^{-1}\RW/\Hm < \alp^{-1/2}$,
the log of its relative fluid potential is
$\Phib\simeq \alp^{-1/4}\gg 1$, $\alp\simeq \rhoi/\rhoo\ll 1$, $\rho$ is the
energy density,
and the subscripts ``in'' and ``out'' refer to quantities inside and outside
the void, respectively.
Because of the time dilation effect, the first-crossing time
is smaller than the outside Hubble time if the void is empty enough.
This effect occurs because the (relativistic or nonrelativistic)
void has a large, negative fluid
potential with respect to the relativistic
background spacetime, which is caused by
relativistic fluid effects.  (In a matter-dominated universe,
these fluid effects are tiny when
the pressure is negligible.)
Because the void is underdense, the time dilation effect is opposite
of that for an overdense region (e.g. a black hole)---time is dilated
in an underdense region and contracted in
an overdense region.
If a superhorizon-sized void is empty enough then, it can collapse in less than
an outside Hubble time.

In this paper,
we study in more detail the collapse of a superhorizon-sized
void embedded in a FRW radiation-dominated universe.
By collapse, we mean the evolution up to the time the shock
(created by the void wall caving in) collides with
itself at the origin.
In particular, we examine how large outward wall velocities affect
the solution and also how gravitational forces affect the solution
at the collapse time.\footnote{Because we find in general
that a void collapses at the speed
of light, we will use the phrases ``collapse time'' and
``first-crossing time'' interchangeably.}
We find that having a large outward wall velocity does not prevent
a superhorizon-sized void
from collapsing in less than an outside Hubble time.
In addition,
we show that if a void collapses in less than an outside Hubble time,
gravitational effects are not important at the collapse time.
This motivates us to compare the collapse of a superhorizon-sized void with
that of a
special relativistic void.  We therefore derive the analytic collapse solutions
in the slab limit, and test our code against these solutions.
At the collapse time, thermalization and homogenization
has not yet occurred, however.
A companion paper\re{Vasha1} will examine the subsequent evolution
of a superhorizon-sized void {\it after} it has collapsed.

The organization of this paper is as follows.
In Section II we present the metric and equations of motion used
to solve this problem.
In Section III,
we describe the improved numerical scheme used
to accurately find the solution near the origin,
and we test this new code on the FRW radiation-dominated universe.
In Section IV, we study the collapse of a superhorizon-sized void both
analytically and numerically.
In Section V we
derive the solution for the collapse of an uncompensated
void in the slab limit.
Finally, Section VI contains a discussion of our results.

\centerline{\bf{II. Spherically Symmetric General Relativistic Fluid
Equations}}
\setcounter{section}{2}
\setcounter{equation}{0}

We are interested in evolving a void embedded
in a FRW universe.  Because we know the initial conditions on an
initial comoving time slice, we choose to work in Lagrangian synchronous
coordinates.
The metric then is
\be
\label{eq:metric}
ds^2=-c^2\Phi^2(t,r)dt^2+\Lam^2(t,r)dr^2+R^2(t,r) d\Omega^2,
\ee
where $t$ is the coordinate time, $r$
is the comoving radius, ~$2\pi R$ is the spacelike circumference
of a sphere centered on the origin,
and $\ln\Phi$ is the
pressure gradient-induced ``potential''.
We consider a perfect fluid with artificial viscosity.
This fluid consists of particles
with mass $\mu$ and temperature $T$.
The stress-energy tensor in this case is
$T^{\alp\bet}=c^{-2}(\rho +p+Q)u^\alp u^\bet+(p+Q)g^{\alpha\beta}$,
where $u^{\alpha}=(-c\Phi^{-1},0,0,0)$ is the fluid $4-$velocity,
$\rho=nc^2(1+\ep/c^2)$ is the energy density, $p=nT/\mu=\w n\ep$
is the pressure,
$n$ is the mass density, $\ep$ is the specific energy, $Q$ is
the artificial viscosity and $\w$ is a constant.\footnote{Boltzmann's
constant has been set to one.}
The relativistic limit is obtained when $\ep/c^2=T/(c^2 \w\mu)\gg 1$.
The viscosity is non-zero only
in shocks.
We assume that the total number of particles per comoving volume is constant,
$\der_\mu(nu^\mu)=0$, so that
\be
\label{eq:defnf}
4\pi nR^2R'/\Gamma=f(r),
\ee
where $~ ' ~\equiv \partial/\partial r$, $f(r)$ is chosen to be $r^2$,
\footnote{This choice only makes sense if
$\Gamma=1$ far outside the void, as will be clear following \eqr{eq:FRWeq}.}
and
\be
\label{eq:Gamlam}
\Gamma \equiv R'/\Lambda.
\ee
The fully general relativistic equations can then be written
as\re{MisSharp},\re{MayWhite}
\ba
\label{eq:dR}
\dR&=&\Phi U \\
\label{eq:dU}
\dU&=&-\Phi\left(\frac{G_NM}{R^2}+\frac{4\pi G_N(p+Q)R}{c^2}\right)-
\frac{\Gamma^2 \Phi (p+Q)'}{{\rm W} n R'} \\
\label{eq:dM}
\dM&=&-4\pi(p+Q)R^2\Phi U/c^2\\
\label{eq:dn}
\dn&=&-\frac{n\Phi(R^2U)'}{R^2R'}\\
\label{eq:dep}
\dep&=&-\frac{\Phi (p+Q)(R^2U)'}{n R^2 R'} \\
\label{eq:defPhi}
\Phi'&=&-\Phi\frac{(p+Q)'}{n{\rm W}c^2},
\ea
where $~\dot{}~\equiv \partial/\partial t$,
\ba
\label{eq:Mprime}
M'&\equiv& 4\pi c^{-2} \rho R^2R'\\
\label{eq:defGamma}
\Gamma^2&\equiv& 1+(U/c)^2-2G_NM/(Rc^2),
\ea
and ${\rm W} \equiv 1+\left[\ep+(p+Q)/n\right]/c^2$
(or $c^2n{\rm W}=\rho+p+Q$).
Here, $U$ is the fluid ``velocity'', $M$ is the ``mass-energy'' and
${\rm W}$ is the relativistic enthalpy.
The energy density equation can be found by combining
Eqs.({\ref{eq:dn}}) and ({\ref{eq:dep}}):
\be
\label{eq:drho}
\dot{\rho}=-\Phi(\rho+p+Q)\frac{(R^2U)'}{R^2R'}.
\ee

\eqr{eq:defGamma} is the conservations of ``energy'' equation, with
total, kinetic and
potential ``energies'' of $c^2(\Gamma^2-1)/2$, $U^2/2$ and
$-G_N M/R$, respectively.
For $G_N=0$, if a particle has velocity $v$ in a Eulerian inertial frame, then
$\Gamma=1/\sqrt{1-(v/c)^2}$ and
$U=\Gamma v$ (see \eqr{eq:Gaminterp}); $\Gamma$ and $U$ represent the
two non-trivial
components of the 4-velocity of the fluid.
We can also write down an important auxiliary equation:
\be
\label{eq:dG}
\dot{\Gamma}=-\Gamma U\Phi\frac{(p+Q)'}{(\rho+p+Q) R'}.
\ee
Thus the ``energy'' $c^2(\Gamma^2-1)/2$ can only decrease in the
presence of pressure gradients.
In SV, \eqr{eq:dG} was used to argue that if $\Gamma\gg 1$
in the void wall, the wall might slow down in a relatively short
amount of time.
We will find later that this is true in some cases.

It is necessary to have artificial viscosity in the code
due to the presence of shocks---it is added
to physically alter the otherwise incorrect solution.
Artificial viscosity dissipates just
enough energy over a shock to satisfy the (exact) jump conditions,
which cannot occur in a perfect fluid.
These conditions
are derived in the appendix, and a demonstration of the
codes ability to give the correct solution with the following
form for $Q$ will
be given there (Figure 13) and later on in this paper (Figure 9).

The form of
the artificial viscosity used here is a generalization of the
nonrelativistic form\re{VonNRich}, and
allows relativistic shocks to penetrate into relativistic fluids.  It does
not, however, alter the solution substantially in non-shock areas.
\ba
\label{eq:Visc}
Q&=&k^2n{\rm W}\Gamma^{-2}(U')^2 dr^2~~~{\rm for~U'<0}\nonumber\\
Q&=&0~~~~~~~~~~~~~~~~~~~~~~~~~~~{\rm otherwise}.
\ea
The constant $k^2$ is of order one, and is proportional to the number
of grid points in the shock.
For relativistic shocks penetrating non-relativistic fluids,
\eqr{eq:Visc} becomes the expression obtained previously \re{MayWhite}.
The expression from this reference however, does not give enough
viscosity over strong shocks when the fluid in front of the shock is
relativistic, because $Q\ll p$.
(It is necessary for the
artificial viscosity to be of order the pressure in the shock region, in order
to dissipate enough energy).
\eqr{eq:Visc} does, however.
Assuming that the fluid in front of the shock is stationary, then
$k^2\Delta U^2\simeq \Gamma^2-1$ over the shock
front, where we have used the shock jump conditions
Eqs.({\ref{eq:Ur}})-({\ref{eq:Gr}}) in the strong shock limit.
Then $Q\simeq (\rho+p)U^2/(\Gamma^2 c^2)\simeq \rho(\Gamma^2-1)/\Gamma^2
\simeq \rho\simeq p$, the desired result.

If the fluid is homogeneous and isotropic,
\eqr{eq:defnf}-({\ref{eq:defGamma}}) reduce to the
FRW equations, with
\ba
\label{eq:FRWeq}
R&=&ra,~~~~
\Phi=1,~~~~
\Gamma=\sqrt{1-\kappa r^2/c^{2}}=\sqrt{1+(RH/c)^2(1-\Omega)},\nonumber\\
M&=&4\pi c^{-2} \rho R^3/3,~~~~
H \equiv \Phi^{-1}\dot{a}/a=U/R,
\ea
where $a(t)$ is the cosmic scale factor,
$H$ is Hubble's ``constant'',
$\Omega=1+\kappa/(Ha)^2=8\pi G_N\rho/(3c^2H^2)$,
$f= r^2/\sqrt{1-\kappa r^2/c^{2}}$,
and $\kappa$ is $-1$, $0$ or $1$ for
negative, zero and positive spatial curvature, respectively.
In a spatially flat ($\kappa=0$) FRW universe then,
$U=c^{-1}R\sqrt{8\pi G_N \rho/3}=\sqrt{2G_NM/R}$.
Note that $\Gamma(t,r)=1$ in a spatially flat universe,
even though the fluid everywhere is moving.

\centerline{\bf{III. Numerical Scheme and Tests of Improved Code}}
\setcounter{section}{3}
\setcounter{equation}{0}

In this section we describe the improved
numerical method used to solve the equations given in the previous section.
We also test the code on the FRW radiation-dominated solution.
Readers more interested in the physical results should skip to section IIIC.
The code is
greatly improved over that used in SV,
because we have now implemented numerical regularization
into the difference equations\re{Evans}.\footnote{It turns
out that scheme 1 from SV for the $\dn$
equation is
differenced correctly, however.}
This is done to get rid of instabilities that develop
at the origin from the inaccurate calculation of
$\dep$, $\dn$ and $\dU_{\rm FLUID}\equiv -\Gamma^2\Phi(p+Q)'/({\rm W}nR')$
there.
As we will see in a moment, the difference equations used in SV
were only accurate to $(\Delta R/R)^2$, where $\Delta R$ is the
grid size, which is or order
one at the first few grid points surrounding the origin.
Thus, no matter how small $\Delta R$ is chosen to be,
$\ep$, $n$ and $U_{\rm FLUID}$
could not be calculated correctly for these grid points.
It is true that decreasing $\Delta R$ confines the inaccuracies
to a much smaller physical volume.  The problem, however,
is that the inaccuracies can become instable
when violent fluid processes take place at or near the origin.  An example
of such a process is when a spherical shock collides at the origin.
These instabilities then cause the code to fail.
Numerical regularization ensures that the error made in calculating
$\ep$, $n$ and $U_{\rm FLUID}$ is of order $(\Delta R/l)^2$,
where $l$ is the characteristic length scale of the problem.
With numerical regularization then, the error made can be a small as you
like by choosing $\Delta R$ sufficiently small.

\centerline{\bf{A.  Numerical Regularization}}

We start by differencing the derivative term in \eqr{eq:dn}
and ({\ref{eq:dep}}), $(R^2U)'/(R^2 R')$, at spatial grid
point $j$ as $\Delta(R^2U)/(R^2\Delta R)$, as was done for the $\dn$ equation
in scheme 2 of SV.
We approximate $R$ and $U$ at adjacent grid points
$j\pm 1$ by Taylor expanding:
\ba
R_{j\pm 1}&=&R_j \pm \Delta r (\partial_r R)_j \nonumber\\
U_{j\pm 1}&=&U_j \pm \Delta r (\partial_r U)_j.
\ea
If we forward difference this derivative term at grid point $j$, we find
\ba
\label{eq:fordiff}
\frac{R^2_{j+1}U_{j+1}-R^2_{j}U_{j}}
{R^2_{j+1}(R_{j+1}-R_{j})}&=&
\frac{1}{R_j^2(\partial_r R)_j}
\left[2 U_jR_j(\partial_r R)_j\left\{1+\frac{\Delta r (\partial_r
R)_j}{2R_j}\right\} \right.\nonumber\\
&+&\left. R_j^2(\partial_r U)_j\left\{1+\frac{2\Delta r (\partial_r R)_j}{R_j}
+ \frac{\Delta r^2}{R_j^2}(\partial_r R)_j^2\right\}\right].
\ea
Backward differencing is obtained by substituting $\Delta r\rightarrow
-\Delta r$ into \eqr{eq:fordiff}.
The forward (backward) differencing gives the predicted (corrected)
solution, as will be described in section B.
The predicted and corrected solutions
are then averaged together.
When this is done, the error made in calculating $\ep$ or $n$
at the new time step is approximately $(\Delta R/R_j)^2$.
This truncation error is small for all but the first few grid points near
the origin.  There, the truncation error is of order one, so that
inaccurate calculation of $\ep$ and $n$ will result
when $U'\neq 0$.\footnote{In SV,
$\ep$ was determined via calculating $\Delta(R^2 U)/(r^2\Delta r)$.
It can be easily
shown that this gives a truncation
error of order ${\cal O}((\Delta R/R_j)^2)$ also.}
If our grid is equally spaced in $R$ so that $R_1=\Delta R/2$,
$R_2=3\Delta R/2$, etc, then for
$j=1$, $2$, $3$ and $4$, the truncation error in the
$R_j^2(\pa_r U)_j$ term is about
400\%, 44\% 16\% and 8\%, respectively.
A similar truncation error problem
occurs when differencing $\dU_{\rm FLUID}$ as in SV:
$\dU_{\rm FLUID}\propto \Delta(p+Q)/\Delta(r^3)$.

We can rectify these problems using numerical regularization.
Near the origin, we expand $p$, $Q$, $R$ and $U$ in a Taylor series
in $r$.  Then we use symmetry arguments to deduce the actual
form of each solution there.
Using the fact that $p(-r)=p(r)$, $Q(-r)=Q(r)$,
$R(-r)=-R(r)$ and $U(-r)=-U(r)$ along with the fact
that $R(0)=U(0)=0$, $p(0)\neq 0$ and $Q_j\propto (U')^2$,
the lowest-order terms in the series at grid point $j$ are
\ba
\label{eq:pQRU}
p_j&=&p_0+p_2 r_j^2 + {\cal O}(r_j^4)\nonumber\\
R_j&=&R_1 r_j + R_3 r_j^3 + {\cal O}(r_j^5)\nonumber\\
U_j&=&U_1 r_j + U_3 r_j^3 + {\cal O}(r_j^5)\nonumber\\
Q_j&=&Q_2 (U_{k+1}-U_k)^2  =
Q_2\left[U_1(r_{k+1}-r_k)+U_3(r_{k+1}^3-r_k^3)
+{\cal O}(r_k^5,r_{k+1}^5) \right]^2,
\ea
where the coefficients $p_0$, $p_2$, etc, are constants, and
where $k=j$ and $k=j-1$ for the forward and backward derivatives,
respectively.
Because $(R^2U)_{j+1}=R_1^2U_1 r_{j+1}^3$ to lowest order, it
is clear that we should
instead difference $(R^2U)'/(R^2 R')$ as
\be
\label{eq:R2Up}
\frac{(R^2 U)'}{R^2R'}=
3\frac{R^2_{j+1}U_{j+1}-R^2_{j}U_{j}}
{R^3_{j+1}-R^3_{j}}
\ee
for the forward derivative, and similarly for the backward derivative.
We can verify this by plugging \eqr{eq:pQRU}
into \eqr{eq:R2Up}:
\ba
3\frac{R^2_{j+1}U_{j+1}-R^2_{j}U_{j}}{R^3_{j+1}-R^3_{j}}&=&
\frac{3 U_1}{R_1}\left[1+(U_3/U_1 + 2R_3/R_1)r_j^2~
\left\{\frac{(1+\Delta r/r_j)^5-1}
{(1+\Delta r/r_j)^3-1}\right\}\right]\nonumber\\
& & \times \left[1+(3R_3/R_1) r_j^2~
\left\{\frac{(1+\Delta r/r_j)^5-1}
{(1+\Delta r/r_j)^3-1}\right\}\right]^{-1}.
\ea
The correct result from \eqr{eq:pQRU} is $3U_1/R_1$.
Since the quantity in curly brackets is of order $4$ or less,
$U_3/U_1\sim {\cal O}((\pa_rR)_j^2l^{-2})$ and
$R_3/R_1\sim {\cal O}((\pa_rR)_j^2 l^{-2})$,
where $l$ is the characteristic length scale of the problem,
the truncation error is of order
$(\Delta R/l)^2$ for $R_j\sim \Delta R$.
Therefore, the error made in calculating this derivative
can be made as small as desired by decreasing the grid size
$\Delta R$.

Now we turn to the derivative term in $\dU_{\rm FLUID}$:
$(p+Q)'/R'$.
Using \eqr{eq:pQRU}, we see that the obvious way to difference
$p'/R'$ is $2R_j(p_{j+1}-p_j)/[(R_{j+1}+R_j)(R_{j+1}-R_j)]$ for the
forward derivative, and similarly for the backward one.
However, this causes a problem at $j=1$ for
the backward derivative
because $p_0=p_1$ and $R_0=-R_1$, but
$R_1(p_1-p_0)/[(R_1+R_0)(R_1-R_0)]\neq 0$.
Instead, we use the simple difference scheme
\be
\label{eq:ppdRp}
\frac{p'}{R'}=\frac{p_{j+1}-p_{j}}{R_{j+1}-R_j}
\ee
for the forward derivative, and similarly for the backward one.
Plugging \eqr{eq:pQRU} into the previous equation, we get
$(2p_2 r_j/R_1) (1+\Delta r/(2 r_j))$ and
$(2p_2 r_j/R_1 )(1-\Delta r/(2 r_j))$
for the forward and backward derivatives, respectively.
Upon averaging
the forward and backward derivatives, we get the correct result,
$2p_2 r_j/R_1$, to lowest
order.\footnote{Although $p_1$ and $R_1$ will be slightly different
for the forward and backward derivatives, their variations are
of higher order.}
Note that in this case, the backward derivative at $j=1$ is zero,
thereby avoiding the problem caused by the former difference scheme.

We also difference the $Q'/R'$ term in the same way as \eqr{eq:ppdRp}
with $p$ replaced by $Q$.
Using $Q_j$ from \eqr{eq:pQRU} with $k=j$, the forward
derivative is
\ba
\frac{Q_{j+1}-Q_{j}}{R_{j+1}-R_j}&=&\frac{12Q_2U_1U_3\Delta r^2 r_j} {R_1}
\left(1+\frac{\Delta r}{r_j}\right) \nonumber\\
&+& \frac{36 Q_2U_3^2\Delta r^2 r_j^3} {R_1}
\left(1+\frac{3\Delta r}{r_j}+
\frac{10}{3}\left(\frac{\Delta r}{r_j}\right)^2
+ \frac{4}{3}\left(\frac{\Delta r}{r_j}\right)^3 \right).
\ea
The backward derivative can be obtained by substituting $-\Delta r$
in for $\Delta r$ in the previous expression.
We can now average the forward and backward expressions and
obtain to lowest order\footnote{Again, we ignore the variations
of $U_1$, $U_3$, $Q_2$ and $R_1$ in the averaging, since
they are of higher order.}
\ba
\left(\frac{Q'}{ R'}\right)_j&\simeq& \frac{12Q_2U_1U_3\Delta r^2 r_j}{R_1}
\left[1+ (3U_3/U_1) r_j^2\left\{ 1+\frac{10}{3}
\left(\frac{\Delta r}{r_j}\right)^2\right\}\right].
\ea
The correct result, from \eqr{eq:pQRU}, is
$12 Q_2U_3\Delta r^2 r_j(U_1+3U_3 r_j^2)/R_1$.
But because $U_3/U_1\sim {\cal O}((\pa_rR)_j^2 l^{-2})$,
where $l$ is the characteristic length scale of the problem,
the quantity in brackets
is $[1+{\cal O}(\Delta R/l)^2]$ for $R_j\sim \Delta R$.
Therefore the error
can be made as small as desired by decreasing the grid size $\Delta R$.

\centerline{\bf{B.  Numerical Procedures}}

We set up a grid with $j_B+1$ grid points
which is equally spaced in $R$ such that
$\dr\equiv R(\ti)_{j+1}-R(\ti)_j$ for $j\in [0,j_B]$, with $R_0=-\dr/2$
and $R_1=\dr/2$.  Then we specify $\rho(R(\ti))$ (or $M(R(\ti))$)
and $\Gamma(R(\ti))$ (or $U(R(\ti))$).
When $\Gamma(\ti,R)$ is given, we always choose to have
the fluid moving outward initially, $U>0$, using \eqr{eq:defGamma}.
Assume that the fluid is initially an
isentrope, so that the specific energy, $\ep$, can be determined via solving
\be
\rho(\ti,R)=\frac{\rho(\ti,\Rmx)}{1+\ep(\ti,\Rmx)/c^2}
\left(\frac{\ep(\ti,R)}{\ep(\ti,\Rmx)}\right)^{1/\w}
\left(1+\ep(\ti,R)/c^2\right)
\ee
iteratively, where the subscript ``$B$'' denotes the quantity at the outer
boundary.  In addition, $\Phi(\ti,R_B)=1$,
and the artificial viscosity is initially
zero: $Q(\ti,R)=0$.  Finally,
$r_j$ and $\Phi(\ti,R)$ can be determined by integrating Eqs.({\ref{eq:defnf}})
and ({\ref{eq:defPhi}}) from $R=0$ and $R=R_B$, respectively.
It is important to note that for
$p=\rho/3$ and $Q=0$,
$\Phi$ can be solved exactly on the initial time slice:
\ba
\label{eq:Phirel}
\Phi(\ti,R)=\Phi_{\rm out}(\ti)
\left(\frac{\rhoo}{\rho(\ti,R)}\right)^{1/4}~~~~~{\rm for~}\ep(\ti,R)/c^2>1,
\ea
where ``out''
refers to quantities outside the void at $j=j_B$.
We define $\rhob$ as the energy density where the fluid becomes
nonrelativistic: $\overline{\ep}/c^2\simeq 1$.
Since the contribution to $\Phi$ is negligible when the fluid
is nonrelativistic,
\ba
\label{eq:Phinrel}
\Phi(\ti,R)=\Phi_{\rm out}(\ti)
\left(\frac{\rhoo}{\rhob(\ti,R)}\right)^{1/4}~~{\rm for~}\rho(\ti,R)<\rhob.
\ea
Inside the void,
$\rhob=\rhoi$ for a relativistic
void, and
$\rhob\simeq g (c^2 \mu)^4$ for a nonrelativistic void,
where $g$ is the number of degrees of freedom,
$\mu$ is the particle mass, and ``in'' refers to quantities inside
the void.
For a nonrelativistic void in a relativistic background fluid then,
\be
\label{eq:PhiNR1}
\Phi_{\rm in}(\ti)\simeq \Phi_{\rm out}(\ti)\ep_{\rm out}(\ti)/c^2.
\ee

We use the MacCormack predictor-corrector method to integrate the equations of
motion.
Suppose we know all quantities on the $i^{th}$ time slice for all spatial
points $j$.
Using \eqr{eq:dn} as an example,
we first predict the new quantities (with forward differencing) for all $j$:
${n_p}^{i+1}_{~j}=n^{i}_j-\Delta t~3n^{i}_j \Phi^{i}_j~
({{R^{i}_{j+1}}^2 ~U^{~i}_{j+1}-{R^{i}_j}^2 ~U^{~i}_{j}})/
({R^{~i}_{j+1}}^3-{R^{~i}_{j}}~ ^3)$.
After using similar methods to obtain ${U_p}^{i+1}_{~j}$, ${R_p}^{i+1}_{~j}$,
${M_p}^{i+1}_{~j}$ and ${\ep_p}^{i+1}_{~j}$,
we integrate \eqr{eq:defPhi} inwards from the outer boundary
using the $4^{th}$-order Runge-Kutta method with linear interpolations
to determine ${\Phi_p}^{i+1}_{~j}$.
The ``predicted'' viscosity (to be used in the corrector step) is:
${Q_p}^{i+1}_{~j}=k^2 n_j^i W_j^i(\Gamma_j^i)^{-2} (U^i_{j+1}-U^i_{j})^2$
from \eqr{eq:Visc}.
We then integrate again (with backward differencing) and
average to obtain the corrected values
${n}^{i+1}_{~j}=.5(n^{i}_j+{n_p}^{i+1}_{~j}
-\Delta t~3{n_p}^{i}_j {\Phi_p}^{i}_j~
({{{R_p}^{i}_{j}}^2 ~{U_p}^{~i}_{j} - {{R_p}^{i}_{j-1}}^2 ~
{U_p}^{~i}_{j-1}})/
({{R_p}^{~i}_{j}}^3-{{R_p}^{~i}_{j-1}}~ ^3)$.
The other quantities are obtained similarly.  (Note that the velocity is
backward differenced also when calculating $Q_{~j}^{i+1}$).
After the $n^{th}$ corrector step, we calculate the time step
for the $(n+1)^{th}$ integration,
$(\Delta t)^{n+1}$,
which must satisfy the Courant condition and the condition that $n$, $\ep$ and
$M$
change slowly enough at each grid point.
(This last condition is necessary for gravitational expansion and contraction).
Thus
\be
\label{eq:calcdt}
(\Delta t)^{n+1} = {\rm min}_j\left(~~C~
\frac{R_{j+1}^n-R_j^n}{\Gamma_j^n\Phi_j^n (c_S)_j^n}~, ~~
\left[\f~\overline{\frac{n_j^n}{|\dn_j^n|}},
{}~~\f~\overline{\frac{\ep_j^n}{|\dep_j^n|}},~~
\f~\overline{\frac{M_j^n}{|\dM_j^n|} }~\right]_{G_N\neq 0}\right),
\ee
where $(c_S)_j^i=\sqrt{{(1+\w)(p_j^i+Q_j^i)}/(n_j^i{\rm W}_j^i)}$
is the speed of sound,\footnote{We include $Q$ in the expression for
$(c_S)_j^i$ because we want the effective pressure.  In practice, it makes
little difference.}
$\f<1$ is a constant, and $n/|\dn|$, etc, are averaged over a few
surrounding grid points.

At the origin, reflecting boundary conditions are used.
And at the outer boundary, $\Phi(t,R_B)=1$,
$p'=n'=\ep'=0$ and $Q(t,R_B)=0$, so that
Eqs.~({\ref{eq:dR}})-({\ref{eq:dM}}) can be easily integrated
using the MacCormack method.
Then $n$ and $\ep$ are determined using the ``free-string''
condition: $n_{j_B}=n_{{j_B}-1}$ and $\ep_{j_B}=\ep_{{j_B}-1}$.

\centerline{\bf{C.  Void Initial Conditions}}

A void is defined by its radius $\RW$, wall thickness $\rw$,
and its relative energy density
$\alpha\simeq \rho_{\rm in}(\ti)/$ $\rho_{\rm out}(\ti) < 1$.
If $c^{-1}\RW/\Hm>1$, the void is said to be superhorizon-sized.
The void wall can be either compensated or uncompensated.
If the void is initially uncompensated, then the
energy density distribution used here is
\be
\label{eq:uncompen}
\rho(\ti,R)=.5\rho_{\rm out}(\ti)[(1+\tanh x)+\alpha(1-\tanh x)],
\ee
where $x\equiv \x$.
Thus the energy density increases monotonically
to its outside FRW value, $\rhoo$.
If the void is initially compensated, then the ``mass-energy''
distribution used here is
\be
\label{eq:compen}
M(\ti,R)=.5c^{-2}4\pi\rho_{\rm out}(\ti)\left[(1+\tanh x)
+\alpha(1-\tanh x)\right]R^3
(\ti)/3.
\ee
In this case, $M$ reaches its
FRW value outside the void.
Therefore, the excess mass-energy density in the void wall compensates
for that missing from the void.
Using \eqr{eq:Mprime}, one finds that the energy density for a compensated
void is
\be
\label{eq:compenrho}
\rho=\rhoo \left\{ \frac{R}{6\rw \cosh^2 x} + \frac{1}{2}
\left[(1+\tanh x)
+\alpha(1-\tanh x)\right]  \right\}.
\ee
The actual
wall thickness can be found by determining the radius, $R_{\rm inner}$,
at which the energy
density is approximately that inside the void.
Setting $\rho(\ti,R_{\rm inner})\simeq  3\rhoi$ and using \eqr{eq:compen}, we
can
solve for $R_{\rm inner}$:
\be
\label{eq:Rinner}
\frac{R_{\rm inner}-\RW}{\rw}\simeq  \ln \sqrt{\frac{3\rw}
{R_{\rm inner}}\alp}
\simeq 1.2\log_{10}\alp + \ln \sqrt{\frac{3\rw}{R_{\rm inner}}},
\ee
where we have assumed that $3\rw\ll \RW$.
Thus the actual wall thickness is larger than
$\rw$, since it is of order or greater than
$|\log_{10}\alp| \rw$.

The initial velocity distribution is determined via
\be
\label{eq:initGam}
\Gamma(\ti,R)=1+(\GW-1)\exp[-(R-\RW)^2/(2\sig_{\Gamma}^2)],
\ee
where $\sig_{\Gamma}$ and $\GW>0$ are
constants.
For this function, $\Gamma$ is equal to one inside and outside the void.
In the void wall, $\Gamma$ can be greater than or equal to one.

In many instances, we would like to match the position of the
inner edge of the energy density distribution
with that of the inner edge
of the velocity distribution.
At $R=R_{\rm inner}$,
the excess kinetic energy should be no larger than the ``potential energy''.
Setting $\Gamma=1+\beta$ with $\beta\ll 1$ and using \eqr{eq:defGamma},
we require
$2c^2\beta\la 2G_N M/R$.  This occurs when
\be
\label{eq:beta}
\beta = \frac{\b}{2} \left(\frac{c^{-1}R_{\rm inner}}{\Hm}\right)^2~\alp,
\ee
where $\b\leq 1$.
Using \eqr{eq:initGam}, the width of the velocity distribution then is
\be
\label{eq:sG}
\sG = \frac{\RW-R_{\rm inner}}
{\sqrt{2\ln[(\GW-1)/\beta]}}.
\ee
For nearly all of the simulations in this paper, we set
$b=1$ to determine $\sG$.
As an example, if $\RW=50$, $\rw=1$,
$\alp=10^{-4}$, $\GW=6$,
$\Hm=2$ and $\b=1$, then $R_{\rm inner}=44$,
$\beta\simeq  .024$, and $\sG\simeq 1.8$.

As a final note, a void from first-order inflation is compensated and
has a large outward wall momentum\re{FOBub}.  The mass ($M$)
outside the void then, is the same as if the void were not present, and
the wall ``velocity'' is large: $\GW >1$.

\centerline{\bf{D.  FRW radiation-dominated model}}

In this section, we test our improved code out on the FRW $\kappa=0$
radiation-dominated model.
We also apply a convergence test since the exact solution is known.
The relative error in $q$, where $q$ denotes any quantity,
is defined to be
$e_i=(q_i-\widetilde{q}(r_i))/\widetilde{q}(r_i)$,
where $\widetilde{q}(r_i)$ is the exact solution and $q_i$ is the numerical
solution.  We obtain a global measure of the error by defining
\be
\label{eq:conv}
L_1=\frac{1}{N}\sum_{i=1}^N |e_i|,
\ee
where $N$ is the total number of grid points.
This error is proportional to the grid spacing to some power:
$L_1\propto \Delta R^s$, where $s$ is the convergence rate.
If $s\simeq 2$, the code is second-order, as desired.
These tools have been used previously to test codes in other
applications\re{Stoneetal}.

For the simulations in this section, we set $\alp=1$,
$G_N=1$, $c=1$, $\w=1/3$,
$\ti=1$, $\ep_{\rm out}(\ti)/c^2=10^6$ and $k^2=0$.
We examine a fluid which is initially relativistic, homogeneous and isotropic,
with
$\Gamma(\ti,R)=1$.
The solution is
$R(t,r)=R(\ti,r)(t/\ti)^{1/2}$,
$4\pi G_N\rho_{\rm hom}(\t)=3c^2/(8t^2)$ and $U=R/(2t)$.

We set
$c^{-1}R_B/H^{-1}(\ti)=250$, $\DR=2.5$ and $C=.3$, and run the code
to $t=1.1$.  These are the same initial conditions used
to make {\it Figure 4} in SV.
{\it Figure 1} shows the relative error in $\rho$ plotted versus the scaled
radius at this
time for $\f=.01$, $.005$ and $.0025$ as solid, dotted and dashed lines,
respectively.
We see that the relative error in $\rho$ is not only independent of $R$, but is
 also very
small, as compared to {\it Figure 4} in SV.
For instance, for $\f=.01$, the inner and outer boundaries have
$-6$\% and $-2$\% errors, respectively, while it is only
$3\times 10^{-4}$\% here.
Thus, the energy density is no longer
underpredicted for the first few grid points near the origin, which
is an important consequence of numerical regularization.

We now calculate the convergence rate in this model.  Since $G_N\neq 0$,
the size of the time step is determined by calculating the smallest of
the following:
$\Delta t_{c_S}\equiv C\Delta R/(\Gamma\Phi c_S)\simeq
\sqrt{3}C\DR\sqrt{t/\ti}$
and $\Delta t_n\equiv \f ~|n/\dn|\simeq \f~2t/(3\ti)$
from \eqr{eq:calcdt}.
Thus we will test our code in two limits:
when $\Delta t_{c_S}<\Delta t_n$ (so that decreasing $\DR$ decreases
$\Delta t$), and
when $\Delta t_n<\Delta t_{c_S}$ (so that decreasing $\f$ decreases $\Delta
t$).
Running our code in the former limit, we set
$R_B=40$, $\f=.5$ and $C=.001$, and run the code
to $t=1.035$.  (Note that we have made $\f$ large and $C$ very small in order
to achieve this limit).
In {\it Table 1}, we show the values of $L_1(q)$ for $q=n,~R,~U,~M$ and $\ep$
for various
$\DR$, and calculate the convergence rate.  We see that the
rate is roughly $2$ for all variables, although there is some variation.

We also run our code in the latter limit, and thus set
$R_B=40$, $\DR=5.0$ and $C=.3$, and run the code to $t=1.014$.
In {\it Table 2}, we show the values of $L_1(q)$ for
various $\f$.
We also calculate the rate $\hat{s}$, defined by
$L_1\propto \f^{\hat{s}}$, in order to determine the speed with which the
code converges when the time step is determined solely
by gravity: $\Delta t_n\propto\f$.
Again, the rate is roughly $2$.

\centerline{\bf{IV.  Collapse of a Superhorizon-sized Void}}
\setcounter{section}{4}
\setcounter{equation}{0}

In SV, we found that the superhorizon-sized voids examined collapsed at the
speed
of light.  The collapse occurs because the fluid in the void wall shocks inward
from the wall pressure and velocity gradients, and it does so at the speed of
light
because the shock is in general strong (see Eqs.~({\ref{eq:shspeedNR}})
and ({\ref{eq:shspeed}})).
In addition, it was found that if the relative energy density in the void is
small enough,
the collapse occurs in less than an outside Hubble time (i.e. a Hubble time
outside the void).
We ran many examples for which the wall energy,
$(\Gamma^2-1)c^2/2$, was zero.
However, we only ran one example ({\it Figure 11}) for which
it was non-zero and large.
The void still collapsed in this case.  At the collapse
time, $\Delta t_{\rm collapse}/\Hm\simeq .7$, there was
substantial pressure and velocity
distortion at the peak and in the peak area.
We would like to understand this example and generalize to other situations.
Do all superhorizon-sized voids
collapse, even if the wall energy is enormous?
What causes the distortion in the peak area, and does
it always occur?
And how does the wall energy affect the collapse time?

In order to answer these questions, we will take two approaches.  First, we
calculate analytically the time scales for change of the fluid in
the void wall as a function of the void radius,
wall thickness, wall velocity and relative energy density, $\alp$.
Second, we answer these questions by showing the results
of many numerical simulations.  We consider different evolutionary
outcomes caused by changing the void's inside energy density,
wall thickness
and excess wall energy.
We then determine
which forces (gravitational or non-gravitational)
are responsible for accelerating the fluid into
the void and for distorting the initial peak and peak area
distributions by comparing runs of
superhorizon-sized voids with those where gravitational forces are neglected.
It turns out that gravitational forces are
negligible if the collapse occurs in much less than
an outside Hubble time.
This may seem confusing,
because superhorizon-sized voids obviously evolve under gravitational forces.
However, the time scales for which gravitational
forces substantially change the solution can be much
larger than that for which fluid forces (gradient forces)
change the solution.
When this is the case, gravitational forces negligibly affect the solution
for some time.
Then, the conclusion that non-gravitational forces have caused virtually
all of the observed changes in the void is a correct
and important result.

\centerline{{\bf A. Time Scales for Fluid Motion in the Void Wall}}

In section B, we will calculate numerically the evolution of many
different superhorizon-sized voids with different initial conditions.
{}From these simulations, detailed evolutionary results can be obtained.
However, it is difficult
to generalize these results to situations with different initial conditions.
For this reason, we
calculate analytically the time scales on which the fluid
in different regions of the void wall changes.
This will allow us to gain intuition over the entire parameter regime
of initial conditions.

Consider a compensated superhorizon-sized void with initial energy density
given
by \eqr{eq:compenrho} with $\alpha\ll 1$ and $3\rw\ll \RW$.\footnote{This
is not a requirement that
the void wall be very thin, because the actual wall
thickness is greater than or of order $|\log_{10}\alp|\rw$ from
\eqr{eq:Rinner}.}
In addition, the initial velocity is
determined via \eqr{eq:initGam}, and the pressure is $p=\w\rho$,
where $\w=1/3$ ($\w=0$) if the fluid is relativistic
(nonrelativistic).
We assume that the fluid outside the void
and in the peak area of the wall is
relativistic, whereas the fluid in the void
may be relativistic or nonrelativistic.
The wall pressure, ``mass'' and ``velocity'' at
the peak, $R=\RW$, is then
\ba
\pW(\ti) &\simeq& p_{\rm out}(\ti)\frac{\RW}{6\rw}\\
M_{\rm w}(\ti) &\simeq& \frac{1}{2} \left(\frac{4\pi\rhoo}{3c^2} \RW^3\right)\\
\label{eq:UW}
U_{\rm w}^2(\ti) &\simeq&
c^2(\GW^2-1)+\frac{1}{2}\left(\frac{\RW}{\Hm}\right)^2.
\ea
Note that the wall energy, $c^2(\GW^2-1)/2$,
is the major contributor to the
wall velocity only if $\GW > c^{-1}\RW/\Hm$.

We now calculate the dynamical time scale on which the fluid
at a particular position changes.  To do this, we first calculate the time
for each of the fluid
variables $R$, $U$, $M$, $\rho$ and $\Gamma$ to change by
assuming that all variables are constant and are given
by their initial values.\footnote{The time
scales of change for $n$ and $\ep$ differ from that for $\rho$
by a number of order one.}
The change in velocity is the sum
of the gravitational ($G_N\neq 0$ and $(p+Q)'=0$) and
``fluid'' ($G_N=0$ and $(p+Q)'\neq 0$) contributions, which we consider
separately.
The fluid variable which changes the fastest then sets the time scale
for change of {\it all} variables.

As an example, we calculate the
time scales for which $\rho$ and $U_{\rm FLUID}$ change.
We set the quantities in Eqs.({\ref{eq:dU}}) and ({\ref{eq:drho}})
equal to their initial values, and calculate the times
at which $\Delta t |\drho |/\rho\simeq 1$
and $\Delta t |\dU_{\rm FLUID} |/U\simeq 1$.
The time scales for change of $\rho$ and $U_{FLUID}$ then are
\ba
\label{eq:Deltall}
\Delta t~[\rho] &\simeq& \frac{\Phi^{-1}}{(1+\w)} \frac{\RW^2 R'}
{(R^2U)'} ;~~~
\Delta t~[U_{FLUID}]\simeq \frac{\Phi^{-1}(1+\w)\rho ~U R'}
{c^2\Gamma^2\w \rho'},
\ea
respectively.
A similar procedure can be followed to obtain
$\Delta t[R]$, $\Delta t[U_{GRAV}]$, $\Delta t[M]$ and $\Delta t[\Gamma]$
from Eqs.({\ref{eq:dR}}), ({\ref{eq:dU}}), ({\ref{eq:dM}})
and ({\ref{eq:dG}}), respectively.
The time scale for change of $\rho$, for example, is given by the
smallest $\Delta t$ obtained, rather than $\Delta t~[\rho]$.
This is because we can no longer assume that the
fluid quantities are equal to their initial values if some other quantity
is changing on faster time scales.  If
$U_{FLUID}$ changes the fastest, for example, $R$ and $U$ (and therefore
$\rho$) will also change on this shorter time scale.

We are interested in the following two regions: the fluid
near the peak ($R=\RW \pm \rw$)
and at the peak ($R=\RW$) of the wall.
When $x=-1$, the energy density
is $\rho=.42\rhoW\ga \rhoo$ and
the spatial derivatives are
$(R^2U)'/(R^2  R')\simeq \UW/\rw$ and
$ \rho'/ R'\simeq \rhoW/\rw$.
If the void is larger than four times
the outside Hubble radius,
then the energy density changes the fastest
\ba
\label{eq:tpeakarea}
\Delta t_{peak ~area}=
\Delta t[\rho]&=& \frac{1}{1+\w}\Phi^{-1}\frac{\rw}{\UW}
=.39 \frac{\left(\RW\rw^3\right)^{1/4}}{\UW}\nonumber\\
&\la&   .55 \Hm \left(\frac{\rw}{\RW}\right)^{3/4},
\ea
where we have used the fact that
$\UW\geq \RW/(\sqrt{2}\Hm)$.
The same time scale for change
is obtained
at $x=1$ as long as $U_{x=1}\ll \UW$.
As the wall gets thinner or the wall energy gets larger, the time scale
for change decreases because of the steeper velocity gradient.

Thus we find the interesting result that
the fluid in the peak area always moves in less than an outside Hubble time.
Note that the time scale for change
is independent of $\alp\simeq \rhoi/\rhoo$, which is not
very surprising because
the velocity gradient will hardly increase
if $\rhoi$ is decreased.

The second region of interest is at the peak ($x=0$), where $\rho'=\Gamma'=0$.
Then $(R^2U)'/(R^2R')\simeq 2\UW/\RW+ (\RW H_{\rm out}(\ti))^2/(4\UW\rw)$, so
that
mass-energy sets the time scale for change:
\ba
\label{eq:tpeak}
\Delta t_{\rm peak}=
\Delta t[M]=3\PhiW^{-1}\frac{\rw}{\UW}&\simeq &
1.9 \frac{(\RW\rw^3)^{1/4}}{\UW}\nonumber\\
&< & 2.7 \Hm\left(\frac{\rw}{\RW}\right)^{3/4},
\ea
which is $5$ times larger than $\Delta t_{\rm peak~area}$.
Again the time scale for change is less than
the outside Hubble time.
Therefore a superhorizon-sized void must collapse in at least less than a
Hubble time if
the fluid in the peak area is to remain virtually stationary
during the collapse.

Everywhere in the void wall, the time scale for change of the energy density is
$\Delta t [\rho]\simeq (1+\w)^{-1}\Phi^{-1} {\rw}/{\UW}$,
which is greater than or equal to the time scale of change at that position.
As we move from the peak area into the void,
$\Phi^{-1}$ decreases
because the fluid ``potential'' decreases due to
the time dilation effect.
Thus the time scale for change decreases as $R$ decreases,
so that the fluid near the base of the void
accelerates into the void on time scales
faster than fluid in the peak area can move; a
superhorizon-sized void cannot prevent its own collapse by having excess
energy in the peak wall area.

We can compare the time scale for change of a superhorizon-sized void
to a void with nearly identical initial
conditions,
but for which the gravitational forces are zero:\footnote{$\RW,~\rw,~n(\ti,R)$,
$\ep(\ti,R),~\Phi(\ti,R),~R(\ti,r)$ and $\Gamma(\ti,R)$ are the same.
Only $U(\ti,R)$ differs through \eqr{eq:defGamma}.} $G_N=0$.
This is a SR void.
At $x=-1$, we assume that $\GW>\Gamma>1$.
Then $\Delta t[U_{FLUID}]$ is the smallest only if
$\Gamma-1\ll 1$ and $\Gamma-1<.1/(\GW^2-1)$.
Otherwise, $\Delta t[\rho]$ is the smallest:
\ba
\label{eq:dpa}
\Delta t_{peak ~area}&=& \Delta t [\rho]
\simeq \frac{1}{(1+\w)} \Phi^{-1}\frac{\rw}{\UW}
={.39} \frac{\left(\RW\rw^3\right)^{1/4}}{\UW}.
\ea
When $U_{x=1}\ll \UW$, the time scale for change is given by
\eqr{eq:dpa} also.
This agrees with the results in the GR case when
$\GW\gg \RW/(c\Hm)$.
When $\GW<\RW/(c\Hm)$, $\Delta t_{\rm peak~area}$ is smaller
in the GR case because $\UW$ is larger.

The time scale for change at the
peak is quite different in the SR case however, because the energy density
now sets the time scale for change:
\be
\label{eq:delSRpd}
\Delta t_{\rm peak}=
\Delta t[\rho]\simeq \frac{3}{8}\Phi_{\rm w}^{-1} \frac{\RW}{\UW}=
.24 \frac{\RW^{5/4}}{\rw^{1/4}\UW}.
\ee
This is larger than the GR result by the factor $\RW/\rw \gg1$.
Thus, it takes much longer for the fluid at the peak to
move for a SR void as
compared to a GR void.
This has interesting consequences.
Because
the energy density of fluid at $x=-1$ decreases
more rapidly than that in the peak,
the pressure gradient
in the upper wall area increases.  This causes $\GW$ to decrease, which
slows the wall down rather quickly.
We will see this in section B1 numerically.

As an example, suppose
we consider a superhorizon-sized void with $\RW=50$, $\rw=1$, $\GW=6$
and $\alpha=10^{-4}$ embedded in a FRW $k=0$ radiation dominated
universe with $\Hm=2$.  Then $\UW=19$ so that
the time for the peak and peak area energy density
to change is $.13\Hm=.27$ and $.028\Hm=.055$, respectively.
If we now consider a SR void with the same parameters and for which
$\Gamma>1.003$, then $\UW=5.9$ so that the time scale
for the peak and peak area energy density to change is
$5.4$ and $.18$, respectively, which are much larger times.

It is interesting to calculate the initial conditions a superhorizon-sized void
must
have in order for its peak area energy density to remain roughly
constant at the collapse time.
{}From \eqr{eq:tpeakarea}, it is clear that the collapse
must occur in less than an outside Hubble time.
{}From \eqr{eq:DeltHGR}, this occurs if $\Phib^{-1}c^{-1}\RW/\Hm <1$,
where $\Phib$ is assumed to be nearly constant.
In this case, the collapse time is
\be
\label{eq:tcoll}
\Delta t_{\rm collapse}\simeq \Phib^{-1}c^{-1}\RW.
\ee
In general, $\Phib$ is larger than its initial,
isentropic value, because the fluid is not isentropic across the
inbound shock
which develops.  Assume that $\Phib=(\rhoo/\rhob)^{p/4}$, where
$\rhob$ is defined after \eqr{eq:Phinrel} and
$p\geq 1$.\footnote{From \eqr{eq:Phinrel}, $p=1$ only in the isentropic
limit.  As an example,
$p=.366/.25=1.46$ for much of the collapse of a
thin-walled uncompensated relativistic void, using
\eqr{eq:Phislab}.}
Letting $\Delta t_{\rm collapse}<\Delta t_{\rm peak~area}$, we find that
\ba
\frac{\rhob}{\rhoo}\la \left(\frac{\rw}{\RW}\right)^{3/p}
\left[ \left(\GW^2-1\right)+\frac{1}{2}\left(\frac{c^{-1}\RW}{\Hm}\right)^2
\right]^{-2/p}.
\ea
Therefore, if $\rhob/\rhoo$ is small enough, then at the first-crossing time
the fluid in the wall will not have moved very much.
It is important to emphasize that if $\rw/\RW\ll 1$
or $\UW\simeq c\GW\gg 1$, then
the collapse must occur in {\it much} less than an outside Hubble time in order
for this to occur.

\centerline{{\bf B. Collapse---Numerical Results}}

In sections B1 and B3, we compare the collapse of a superhorizon-sized void
with that of a void
with similar initial conditions, but for which we neglect
gravity.
(This latter void is a special relativistic (SR) void).
The purpose of this is to compare which evolutionary changes in the void
are due to gravitational and non-gravitational
(i.e. fluid) forces, since any effects which appear similarly for both SR and
GR voids
cannot be due to gravitational forces.
We find that if a superhorizon-sized void collapses in less than the
time scale for change in the peak area,
the solution can be approximated very well
by the solution to a SR void with the same initial
conditions.
For these voids then, gravitational effects are not important
during the collapse.
In addition, we find that even if
the collapse time is {\it larger} than the time scale for change in
the peak area, as long as the collapse
occurs in much less than an outside Hubble time, gravitational effects are
unimportant.
These numerical results are important, since they show that the
gravitational contribution to the solution is negligible at the
collapse time if
a superhorizon-sized void collapses in much less than an outside Hubble time,
regardless of the details in the wall area (e.g. wall thickness,
wall energy, etc).

\vspace{.1 in}

For {\it Figures 2-7} and {\it Tables 3-4},
we choose $\Gamma(\ti,R)$ and $\rho(\ti,R)$
from Eqs.({\ref{eq:initGam}}) and ({\ref{eq:compenrho}}), respectively.
We then integrate the equations of motion twice,
starting with the same initial conditions.\footnote{The only
variable which differs between the
GR and SR cases initially is the velocity
$U(\ti,R)$.  All other variables are identical.}
The first time we set $G_N=1$ (the ``GR'' case), which gives the solution
for the evolution of a superhorizon-sized void.
The second time we set $G_N=0$ (the ``SR'' case), which gives
the solution for the evolution of a special relativistic void.
The pressure in the GR (SR) case at time $t$ is defined to be
$p_{GR}(t,R_i)$ ($p_{SR}(t,R_i)$), where $R_i\equiv R(\ti,r)$.

In all figures which follow in section IVB, we set
$C=.3$, $c=1$, $\w=1/3$, $\f=.01$, $\ti=1$,
$\ep_{\rm out}(\ti)/c^2=10^6$
and $4\pi\rho_{\rm out}(\ti)=3/8$.  Thus the initial Hubble time outside
the void is $H^{-1}_{\rm out}(\ti)=2\ti=2$,
and all voids are relativistic.\footnote{A relativistic (non-relativistic)
void is when the fluid in the void is relativistic (non-relativistic).}

\centerline{\bf 1.  Gravitational Effects during the Collapse of a
Superhorizon-sized Void}

We start with compensated voids for which $\RW=50$, $\rw=1$,
$\Gamma(\ti,R)=1$, and $\DR=.25$.
Note that the GR void is superhorizon-sized, since $c^{-1}\RW/\Hm=25$.
For $\alpha =10^{-4}$, $10^{-7}$ and $10^{-10}$,
we set $k^2=1.5,1.7$ and $1.7$, respectively, and
we run our code twice to
$t=4.1$, $t=1.5$, and $t=1.08$, respectively.

In {\it Figure 2}, we show
the fractional difference in the pressure,
$[p_{\rm GR}(t,R_i)-p_{\rm SR}(t,R_i)]/$ $p_{\rm GR}(t,R_i)$,
as a function of the initial radius.
For $\alpha =10^{-4}$, $10^{-7}$ and $10^{-10}$,
we plot solid triangles, open boxes, and a dashed line, respectively.
As the inside energy
density decreases, the difference
between the SR and GR solutions
decreases also.  For $\alpha=10^{-4}$, $10^{-7}$ and $10^{-10}$,
the absolute value of this difference is at most 350\%, 110\%
and 20\%, respectively.
A similar result is obtained for an initially uncompensated void.

When $\alp=10^{-4}$, the solutions at $t=4.1$ are very different.
Most of the fluid not directly behind the shock
is moving {\it outward} in the GR case, as opposed to the SR case.
In addition, the large void wall pressure, $\pW$,
has redshifted away in the GR case, causing
the pressure gradient to be almost nonexistent in the former wall region.
This is in contrast to the SR case, which has hardly changed at all
in the peak area.
The fact that the solution at
$t=4.1$ has changed considerably
in the GR case is not very surprising, because the time scales for change
in the peak and peak area are
$.29$ and $.059$, respectively, from Eqs.({\ref{eq:tpeak}}) and
({\ref{eq:tpeakarea}}).

In {\it Figure 3}, we show voids identical to those of {\it Figure 2} but for
which
$\GW=6$ and $\sG=\rw$.
For this value of $\GW$ however, $\UW/c=19>\GW$, so that the velocity
is about $5$\% higher than for {\it Figure 2}.
Again, as the GR void becomes emptier,
the solution looks more like the SR solution.
For $\alp=10^{-10}$, gravitational effects can be neglected during
the collapse of the superhorizon-sized void.
Note that the relative pressure difference for $\alp=10^{-4}$ is only
$50$\% in the shock area as opposed to $300$\% in {\it Figure 2}.
This is because the SR shock here is weaker since more wall fluid
is pushed out with the larger wall velocity.
A similar result is again obtained for the uncompensated case.

In {\it Figure 4}, we plot the radii of seven comoving
observers as a function of time for the voids in {\it Figure 3} with
$\alp=10^{-4}$ and $\alp=10^{-10}$.
The solid and dashed lines represent the
GR and SR results, respectively.
The figures show that when $\alp=10^{-4}$,
the SR and GR results deviate quickly.
Note that the
comoving observers move outward in the wall region in the GR case.
However, when $\alp=10^{-10}$, the SR and GR results are quite
similar for the entire collapse.
In addition, the peak area looks virtually unchanged in both
cases.  This is easy to explain.
Using \eqr{eq:tpeak} and ({\ref{eq:delSRpd}}),
the peak changes on time scales $.27$ and $5.4$ for
the GR and SR cases, respectively.
And in the peak area, the time scales of change for the GR and
SR voids are $.055$ and $.18$, respectively, using
Eqs.({\ref{eq:tpeakarea}}) and ({\ref{eq:dpa}}).

In {\it Figure 5} we plot
$\rho$ and $\Gamma$ as a function
of time for comoving observers in the peak
area.  These voids
are identical to those of {\it Figure 3} with $\alp=10^{-4}$.
Again, the solid and dashed
lines show the results in the GR and SR cases, respectively.
Note that the energy density
for the GR void in the peak and peak area has decreased
significantly by $t\simeq 1.1$, in agreement with the time
scale $.055$ calculated above.
In addition, $\rho$ keeps decreasing for all three observers.
In the SR case however, the pressure at the peak
hardly changes, while that in the peak region has decreased substantially
by $t\simeq 1.3$, in agreement with the time scale $.18$ calculated above.
In addition, in the SR case the energy
of the peak, $c^2(\GW(t)^2-1)/2$, is substantially less by $t\sim 2$.
In the GR case however, $\GW(t)$ starts to decrease, then
stops when the peak energy density decreases substantially.
Therefore, if a
superhorizon-sized void collapses in more than the outside
Hubble time, the wall energy probably does
not decrease, as it does in the SR case.

\centerline{\bf 2. Collapse Times for Superhorizon-sized Voids}

We now calculate the collapse times for many different superhorizon-sized
voids.
These times depend
on the initial wall energy, $c^2(\GW^2-1)/2$, the relative
energy density, $\alp$, and the wall thickness $\rw$.
Again, we choose a void which is $25$ times the Hubble radius, with
relative wall thickness
$\rw/\RW=1/50$.
We set $\DR=.25$, $k^2=1.7$,
and determine
$\sG$ from Eqs.({\ref{eq:sG}}), ({\ref{eq:Rinner}}) and ({\ref{eq:beta}})
by setting $\b=1$.
We present our results in {\it Table 3}.
The stars following some of the collapse times mark those
voids where there is significant distortion observed in the peak area just
before
the collapse.
It is interesting to compare this with the time
scales for change in the peak area.
Using Eqs.({\ref{eq:tpeakarea}}) and ({\ref{eq:UW}}), the time scales for
change
are given also.
As expected, for a given value of $\GW$, the peak area is distorted
unless the collapse time is smaller than the time scale for change in the
peak area.
In addition, if the peak is distorted and $\GW$ is increased
further, the collapse time increases.
For $\alp=10^{-7}$, $\alp=10^{-7}$ and $\alp=10^{-10}$,
the collapse time increases by
$21$\%, $4$\% and $<1$\%, respectively, for $\GW=6$ to $50$.
This is because more fluid moves out of the peak area (as a shock)
rather than into the void,
weakening the shock.  (The outgoing shocks can be observed
in {\it Figures 6-7}).  This increases the shock
formation time, and decreases the relative potential
of the void because
$\Phi_{\rm in}$ decreases slightly when an outgoing shock is
present also.  This can be seen by rewriting \eqr{eq:defPhi} as
\be
\label{eq:Phivis}
\Phi_{\rm in}(t)=\left(\frac{\rho_{\rm out}}{\rho_{\rm in}}\right)^{\w/(\w+1)}
\exp\left(\frac{1}{\w+1}\int_0^\infty\frac{Q'dr}{\gamma\rho+Q}\right).
\ee
$Q$ is non-zero only in a shock, and is peaked at the center of it.
Divide the shock front into two portions separated by the position where $Q'=0$
and $p=p_Q\simeq Q_Q$.
Since $p\ga Q$ in the shock, for an inbound shock
the contribution to the integral is larger when $p<p_Q$ ($Q'>0$)
than when $p>p_Q$ ($Q'<0$).  Therefore
when the shock is inbound (outgoing), $\Phi_{\rm in}$ increases
(decreases) from its isentropic value.
As an example, if we assume that both inbound and outgoing
shocks are strong, using Eqs.({\ref{eq:phibr}}) and ({\ref{eq:Phirel}})
we find that the relative potential of the void is
\be
\Phi_{\rm in}=\alpp^{-.25}\left(\frac{\rho_{b,i}}{\rho_{\rm in}}\right)^{.25}
\left(\frac{\rho_{\rm out}} {\rho_{b,o}}\right)^{.25},
\ee
where $\rho_{b,i}$ ($\rho_{b,o}$) are the energy densities behind the inbound
and outgoing shocks, respectively, and
$\alpp\equiv \rhoi/\rhoo$.  Thus, the outgoing shock partially cancels
some of the void's relative potential.

In the last two rows,
we calculate the theoretical collapse
times for each $\alp$ using \eqr{eq:tcoll}.
Note that $\Phi_{\rm in}=\alpp^{-.25}$ is the isentropic result, and
$\Phi_{\rm in}=.6 \alpp^{-.366}$ is the relativistic shock tube result
from \eqr{eq:Phislab}.  Therefore,
the numerical
collapse times are nearly that for the isentropic ``no-shock'' result.
This is because the
void wall is too thick, which results in
the production of weaker shocks, large
shock formation times and a smaller time dilation effect.

We can show this by calculating the collapse times
as a function of the width $\rw$
for a superhorizon-sized void
with $\RW=50$, $\alp=10^{-10}$, $\GW=6$, $\DR=\rw/4$ and $b=1$.
For $\rw=1,~.5$ and $.25$, we set $k^2=1.7,~2.2$ and $3.0$, respectively.
The results are presented in {\it Table 4}.  In the last column,
we calculate the exponent $\pp$
using \eqr{eq:tcoll} with the assumption that
$\Phi_{\rm in}=.6\alpp^{-\pp}$.
As $\rw$ decreases, the collapse time decreases, and $\pp$ increases,
as expected.

As a final remark,
because the time scale for change in the peak area of a SR void is
greater than that for a GR void with the same initial conditions,
if a GR void collapses in much less than the time scale for change in the
peak area, we
expect the GR and SR voids to look the same and be unchanged
in the peak area
at the collapse time.\footnote{It is possible to find initial conditions
such that the collapse time is between the GR and SR time scales
for change in the peak area.
However, in order for this to happen for collapse times much less
than the Hubble time, $\rw/\RW$ must be extremely small.  For example,
using \eqr{eq:tpeakarea}, if $\Delta t_{\rm peak~area}/\Hm\simeq .001$
and $.005$, we require $\rw/\RW\simeq 2.3\times 10^{-4}$ and
$1.9\times 10^{-3}$.  These voids have extremely thin walls,
and will not be considered here.}
This is observed numerically
for the voids listed in {\it Table 3}.
The fluid which {\it is} in motion at the first-crossing time
has been accelerated into the void by fluid forces.
Therefore, it is not surprising that
the solutions for these GR voids are nearly identical to those
of SR voids evolved under the same initial conditions.
Thus gravitational effects are not important
at the first-crossing times for superhorizon-sized voids with
$\Delta t_{\rm collapse}\ll \Delta t_{\rm peak~area}$.

\centerline{\bf 3. Peak Distortions: Fluid or Gravitational effects}

{}From the above discussion, it is clear that the wall of a void
will be distorted at the collapse time if $\GW$ is large enough,
even if the collapse occurs in less than an outside Hubble time.
It is therefore important to find out if this distortion is
caused by non-gravitational (fluid) or gravitational effects.
Thus, we compare the GR solution before collapse with that of
the SR solution with the same initial conditions.
We choose the initial conditions such that
$\Delta t_{\rm peak~area}<\Delta t_{\rm collapse}$.

We evolve a GR and a SR void with $\RW=50$, $\rw=1$,
$\alp=10^{-10}$, $\GW=200\simeq \UW/c$ and $\sG=1.88\rw$.
In addition, $k^2=1.7$ and $\DR=.25$.
It is found that
the collapse time for the SR void is
$\Delta t_c=.18$, which is $31$\% larger than the GR collapse time
of $.13$.
In {\it Figure 6},
we plot $p$ and $\Gamma$ as a function of $R(t,r)$
for the GR void (dotted line) and SR void (dashed line)
at $t=1.06$.
The initial void configuration is shown as the solid line.
Note that the GR void's shock is ahead of the SR void's shock
($24<R<26$), which explains why the SR collapse time is longer.
Since the time scale for change in the peak area, $.0052$,
is much smaller than the collapse time, we expect the
large distortions seen in the peak area ($45<R<60$); the
peak has moved outward and is thinner.
In addition, the peak energy is smaller, which is
due partly to the spherical damping effect of the outbound
wave.  However, the important and surprising
result is that the
GR and SR voids look very similar at this time.
It is true that the energy in the void wall should
drive the fluid out regardless of whether
gravitational forces are included or not.
However once the fluid starts expanding outward,
one can imagine gravitational effects adding or
subtracting substantially to this movement.
Apparently, this does not happen.

We would like to know how robust this result is.
Therefore we examine voids with larger and smaller
collapse times and for which peak distortions are large.
We also require $\GW \gg \RW/(\sqrt{2}c\Hm)\simeq 18$ so that the initial
velocity
in the wall is the same for both GR and SR
cases: $\UW\simeq c\GW$.\footnote{If $\GW\ll\RW/(\sqrt{2}c\Hm)$, then
the wall velocity $\UW$ is much larger in the GR case as compared to the SR
case, causing
the time scale for change to be smaller.}
If we choose smaller values for $\GW$, then it is not possible
to distinguish the
non-gravitational from the gravitational effects
in this manner.

For a void with $\RW=50$, $\rw=1$, $\alp=10^{-7}$,
$\GW=50$, $\sG=1.73\rw$, $k^2=1.7$ and $\DR=.25$,
the GR and SR solutions are very different at $t=1.4$.
However, a substantial fraction of an outside Hubble time has passed.
We therefore choose a void which collapses much more quickly and
for which $\GW$ is very large.

We choose a void with $\RW=50$, $\rw=1$, $\alp=10^{-13}$,
$\GW=500$, $\sG=2.05\rw$, $k^2=1.7$ and $\DR=.25$.
{\it Figure 7} shows that the solutions
are nearly identical at $t=1.01$, even though the peak
is fairly distorted.
In addition, the collapse times for the GR and SR void are
$\Delta t_c=.019$ and $.021$, respectively,
which are not very different.
Thus, the gravitational contribution to the solution is
smaller than that for
the superhorizon-sized void in {\it Figure 6}, even
though the peak distortions are comparable.
This is because the collapse time is even smaller.

In conclusion, if a superhorizon-sized void collapses in {\it much} less than
an
outside Hubble time,
gravitational forces can be neglected in the equations of motion
up until the collapse time.
This is true even if the fluid in the peak area moves substantially
during this time.

\centerline{\bf{V. Lagrangian Shock Tube as an Approximate Solution}}
\centerline{\bf{for the Collapse of a Superhorizon-sized Void}}
\setcounter{section}{5}
\setcounter{equation}{0}

We have found that if a superhorizon-sized void collapses in much less than an
outside Hubble time, then even
if the void wall moves outward significantly,
the solution
at the collapse time looks nearly identical to that for a SR void
evolved under similar conditions.
This is an important result, since it shows that the force moving
the fluid out of the peak area is not due to
gravity, but is due to pressure and velocity
gradients which are present in both cases.
(The force moving the fluid into the void is non-gravitational also,
for the same reason).
More importantly, it shows that gravitational forces are not important
until after the void collapses.
This leads us to search for collapse solutions to the special relativistic
equations (i.e. $G_N=0$).

Consider an uncompensated void with a small wall velocity.
As the void starts to collapse, the solution is
approximately slab (linearly) symmetric.  Not until the shock
nears the origin will spherical
effects come into play.  If in addition the wall is thin
($|\log_{10}\alp|\rw/\RW\ll 1$), then it acts like an initial
discontinuity in pressure, energy density, etc.  The solution
for this is called a shock tube.
When the solution is assumed to be spherically symmetric
with analogous initial conditions (i.e. the problem we would like
to solve), a
similarity solution does not exist.
This is because another length scale, the void radius, $\RW$,
has been introduced
into the problem\re{Sedov}.
Thus, the slab shock tube solution derived here will only be important for
uncompensated voids with thin walls, small wall velocities, and
for which $\Delta t_{\rm collapse}\ll \Hm$.

The general relativistic equations used here
do admit similarity solutions\re{CahTaub}.
We will not solve these equations here however,
because we cannot find similarity solutions for our problem.
In Appendix A, we review the conditions satisfied
over a contact discontinuity and shock for a general relativistic
fluid\re{MayWhite}.
In section VA,
we derive the slab similarity solution for relativistic fluids
in this Lagrangian coordinate system.
We then attach this solution to
the contact discontinuity and shock solutions, which are also
similarity solutions, when the fluid in front of the shock
is relativistic and nonrelativistic.
This corresponds to the fluid
in the ``void'' being relativistic and nonrelativistic, respectively.
The relativistic (shock tube) solution was
derived in a Eulerian coordinate system, which can be obtained from
the Lagrangian description by a spacetime coordinate
transformation\re{ShockTube}.
Finally, we compare these solutions to the numerical solutions obtained
from the collapse
of relativistic or
nonrelativistic voids.

\centerline{{\bf A. Relativistic Similarity Solution}}

In this section, we derive the similarity solution for a relativistic fluid
(with no viscosity) in the slab symmetric
limit.  This limit
corresponds to the evolution of a spherically symmetric fluid
distribution which is
``far'' from the origin.  Mathematically, we require
$|U/R|\ll |U'/R'|$ to be satisfied.

We define $X(t,x)=
R(t,r)$ to be the location of a comoving shell with label $x=r$.
In order that the solution be a similarity solution,
all functions can only depend on the variable
\be
z\equiv \frac{x}{t},
\ee
where $t=0$ is the initial time.
Thus, $\rho=\rho(z)$, $p=p(z)$, $U=U(z)$, $\Phi=\Phi(z)$ and
$\chi(z)\equiv X/x$.
($X=R$ and $x=r$ in the slab limit).
We can then rewrite
Eqs.({\ref{eq:dR}})-({\ref{eq:dU}}), ({\ref{eq:dn}}),
({\ref{eq:defPhi}}) and ({\ref{eq:drho}}) as
\ba
\label{eq:drhoz}
z\frac{d\rho}{dz}&=&\frac{\Phi(\rho+p)}{d\chit/dz}\frac{dU}{dz}\\
\label{eq:PhiU}
\Phi U&=&-z\frac{d\chit}{dz}+\chit\\
\label{eq:dUz}
z\frac{dU}{dz}&=&\frac{c^2\Gamma^2\Phi}{(\rho+p)d\chit/dz}~\frac{dp}{dz}\\
\label{eq:dPhiz}
\frac{d\Phi}{dz}&=&-\frac{\Phi}{\rho+p}~\frac{dp}{dz}\\
\label{eq:dnz}
z\frac{dn}{dz}&=&\frac{\Phi n}{d\chit/dz}~\frac{dU}{dz},
\ea
where $\chit\equiv z\chi=X/t$ and
$\ep=\rho/n- c^2$.
We now assume that $\ep/c^2\gg 1$ and that
the equation of state is $p=\w\rho$.
Eqs.({\ref{eq:drhoz}})-({\ref{eq:dnz}}) can then be solved to obtain
\ba
\label{eq:PU}
\Phi U&=&\mp c_S\Phi\Gamma+\chit\\
\label{eq:rhosim}
\rho/\rho_0&=&\left({z}/{z_0}\right)^{(1+\w)/(1-\w)}
=(t_0/t)^{(1+\w)/(1-\w)}\\
\label{eq:Usim}
U/c&=&\frac{(\rho/\rho_0)^{2\beta}-1}{2(\rho/\rho_0)^{\beta}}\\
\Phi&=&\Phi_0(\rho_0/\rho)^{\w/(1+\w)}\\
\label{eq:nsim}
n&=&n_0(\rho/\rho_0)^{1/(1+\w)}
\ea
and $\ep=\rho/n$,
where $\beta\equiv \pm\sqrt{\w}/(1+\w)$, $c_S\equiv \sqrt{\w} c$
is the speed of sound, and where the subscript ``$0$'' denotes the value
a shell has upon entering the region where this solution holds.
Eqs.({\ref{eq:PU}})-({\ref{eq:nsim}}) describe a rarefaction wave going
to the right (left) when the
upper (lower) sign is chosen.
In addition, the position of a shell is given by
\ba
\label{eq:Xrare}
\frac{X(t,x)}{t}=\Phi\left[\frac{\pm2 c_S-(c \pm c_S)
\left\{1-(\rho/\rho_0)^{2\beta}\right\} } {2(\rho/\rho_0)^\beta}\right].
\ea
Therefore, we have solved for all functions in the rarefaction
wave in terms of
$(\rho/\rho_0)$.

\centerline{{\bf B. Shock Tube Solution}}

In this section, we find the solution for a special relativistic fluid with an
initial
discontinuity in the pressure.  The side with the larger pressure
is relativistic, and the fluid on the other side is relativistic
or nonrelativistic.
These solutions will be used to model the collapse of
superhorizon-sized voids in the early universe.

Suppose at $t=\ti$, the fluid is at rest everywhere, with
an initial discontinuity
at position $\XW$.
On either side of the discontinuity, $\rho$, $n$, $p$ and $\ep$
are constant.
The subscripts $1$ and $2$ represent the fluid quantities
for $X(\ti)<\XW$ and $X(\ti) > \XW$, respectively.
We choose the coordinate system where $\rho_2>\rho_1$,
$\ep_2>\ep_1$ and $n_2>n_1$.  The equation of state is
$p=\w\rho$ in region 2, and
$p=\w \rho$ ($p=0$) in region $1$ if the fluid there
is relativistic (non-relativistic).
For $t>\ti$, a shock will be formed
which travels into region 1 (undisturbed fluid), a rarefaction wave will be
formed which moves into region 2 (also undisturbed),
and a contact discontinuity will form in between.
There will be
5 distinct regions in all.
The shock is located at $X_d(t,x_{\rm sh})$,
region 3 contains the fluid behind the shock,
the contact discontinuity is located
at $X_c(t,x)$, and the
rarefaction wave is located in region $4$ for
$X_b(t,x)<X< X_a(t,x)$.
{\it Figure 8} is a sketch of pressure versus distance for a relativistic shock
tube, with the regions and boundaries labeled.

We will assume
that the fluid in the rarefaction wave is relativistic
regardless of the equation of state in region 1, so that
$\ep_{3'}/c^2\gg 1$.  Doing this allows us to
use the similarity solution obtained in VA for the rarefaction wave.
And because $\ep_3>\ep_{3'}$, as we will see in a moment,
it therefore follows that $\ep_3/c^2\gg 1$ in both cases.
Then, the solution in region 3 is given by
Eqs.({\ref{eq:UbNR}})-({\ref{eq:shspeedNR}}) if region 1 is
nonrelativistic, and Eqs.({\ref{eq:Ur}})-({\ref{eq:shspeed}})
if it is relativistic.  In either case, the upper signs
are used because the shock moves in the direction
of decreasing $X$.
Note that these are all given in terms
of $\rho_3=\rho_b$.
The conditions across the contact discontinuity
give $\Phi_3=\Phi_{3'}$, $ p_3=p_{3'}$, $ \rho_3=\rho_{3'}$
and $ U_3=U_{3'}$ from \eqr{eq:contact}.
Finally, the rarefaction
wave solution is given by Eqs.({\ref{eq:rhosim}})-({\ref{eq:nsim}}) with
the upper signs and $\rho_2\equiv\rho_0$.

\centerline{\bf{1.  Relativistic, Thin-Walled Uncompensated Void}}

We first examine the case where the fluid in region 1
is relativistic.
Since $U_{3'}/c=((\rho_3/\rho_2)^{2\beta}-1)/[2(\rho_3/\rho_2)^\beta]$
from \eqr{eq:Usim},
we can use this and \eqr{eq:Ur} to find the equation relating
$\rhob\equiv \rho_3/\rho_1$, $\beta=\sqrt{\w}/(1+\w)$ and
$\delta\equiv(\rho_1/\rho_2)^\beta$:
\be
2\delta\beta(\rhob-1)\rhob^{\beta-1/2}=1-\delta^2\rhob^{2\beta}.
\ee
An iterative method is then used to find $\rhob$.
The solution in region 3, 3' and 4 is given by
Eqs.({\ref{eq:Ur}})-({\ref{eq:shspeed}}),
Eqs.({\ref{eq:Usim}})-({\ref{eq:nsim}}) with $\rho=\rho_3$
and $\rho_0=\rho_2$,
and Eqs.({\ref{eq:rhosim}})-({\ref{eq:nsim}}), respectively.

We first calculate the location of the boundaries between the
five regions: $X_a$, $X_b$, $X_c$ and $X_d$.  Set
$\Phi_2=1$.
We can calculate $X_a$ by noting that the velocity is zero there
and by using \eqr{eq:PU}.
$X_b$ is found by using \eqr{eq:PU} and ({\ref{eq:Ur}}).
Since $U(z)$ and $\Phi(z)$ are constant across the contact discontinuity
and in regions 3 and 3',
we can integrate \eqr{eq:PhiU} to obtain the position of a shell in region
3 or 3':
\be
\label{eq:X33p}
X(t,x)=\Phi U(t-t_0)+X_0.
\ee
Here $X_0$ is the location of a shell when it enters region
3 or 3' at time $t_0$.
\eqr{eq:X33p} also applies to the location of the shock and contact
discontinuity,
with $X_0=\XW$ and $t_0=\ti$.  The locations of the boundaries are then
\ba
\label{eq:Xa}
X_a(t)&=&\XW+c_S (t-\ti) \\
\label{eq:Xb}
X_b(t)&=&\XW+(t-\ti)\Phi_3 \beta\frac{\rhob-1}{\sqrt{\rhob}}
\left[-c+c_S\sqrt{1+\frac{\rhob}{\beta^2(\rhob-1)^2}}\right] \\
\label{eq:Xc}
X_c(t)&=&\XW+(t-\ti)\Phi_3 U_3\\
X_d(t)&=&\XW+(t-\ti)\Phi_3 U_{\rm sh}=\XW+(t-\ti)(1+\w)
\frac{\rhob}{\rhob-1}\Phi_3 U_3,
\ea
where the shock speed is $\Phi_3 U_{\rm sh}\equiv R'_{\rm sh}{\dot r}_{\rm sh}
=X'_{\rm sh}{\dot x}_{\rm sh}$
from \eqr{eq:shspeed}.
Note that the rarefaction wave moves out at the speed of sound.
In addition, from the discussion following \eqr{eq:shspeed}, if
the shock is strong (i.e. $\rhob\gg 1$), the shock moves into region 1 at the
speed of light:
$X_d(t)=\XW-c(t-\ti)\Phi_1$.

Now we trace the location of all Lagrange points in the shock tube
in terms of $z=x/(t-\ti)$.
A shell with coordinate $x$ and position $X$
initially in region 2
($X>\XW$) will enter the rarefaction wave at time
$t_a=\ti+(X-\XW)/c_S$.  Using \eqr{eq:Xrare} and setting
$z_a=x/(t_a-\ti)$,
its position as a function of time in that
wave for $z\leq z_a$ will be
\be
\label{eq:Xrx}
X_4(t,x)=\XW+(t-\ti)\left(\frac{z}{z_a}\right)^
{-\frac{(\w+\sqrt{\w})}{(1-\w)}}\left[c_S
-\frac{c_S+c}{2}\left(1-\left(\frac{z}{z_a}\right)^
{\frac{2\sqrt{\w}}{(1-\w)}}\right)\right].
\ee
Here, the subscript ``4'' shows that the shell is in region 4.
The energy density of this shell decreases until it reaches $\rho_3$,
at which point it enters region 3'.  This occurs at time
$t_b=\ti+(t_a-\ti)(\rho_2/\rho_3)^{(1-\w)/(1+\w)}$.  Its position at this point
is
$X_b(t_b,x)$ and is
given by \eqr{eq:Xb}
or \eqr{eq:Xrx} with $t=t_b$ and $z=z_b=x/(t_b-\ti)$.
Using \eqr{eq:PU}, its position in region 3' for $z\leq z_b$ is
\be
X_{3'}(t,x)=\XW+(t-\ti)\left\{\Phi_3 U_3-\frac{z}{z_b}\left(\Phi_3 U_3-
\frac{ X_b-\XW}
{t_b-\ti}\right)\right\}.
\ee
This Lagrange shell can never cross into region 3 because the contact
discontinuity
moves with the fluid.

Finally, if a shell initially has the position $X<\XW$, then it will enter
region 3
when the shock reaches it.
This occurs at time $t_d=\ti+(X-\XW)/(\Phi_3 U_{\rm sh})$.
For $t> t_d$, the location of this shell in region 3 for $|z|\leq |z_d|$ is
\be
\label{eq:X3}
X_3(t,x)=\XW+(t-\ti)\left\{\Phi_3U_3-\frac{z}{z_d}\left(\Phi_3 U_3-
\frac{X_d-\XW}{t_d-\ti}\right)\right\},
\ee
where $z_d=x/(t_d-\ti)$.

In the strong shock limit ($\rhob\gg 1$), $\rhob$ is given
by an analytic
expression.
The solution then is
\ba
\rho_3&=&\rho_1 ~a^{2(1+\w)}
\alpp^{-2\sqrt{\w}/(1+\sqrt{\w})^2}
\rightarrow 1.2\rho_1 \alpp^{-.464}\\
\frac{U_3}{c}&=&-\frac{\sqrt{\w}}{1+\w} ~a^{(1+\w)}
\alpp^{-\sqrt{\w}/(1+\sqrt{\w})^2}
\rightarrow -.47 \alpp^{-.232}\\
\label{eq:Phislab}
\Phi_1&=&\Phi_2\sqrt{\w} ~a^{(1-\w)}
\alpp^{{\frac{-w}{1+\w}
\left[1+\frac{1-\w}{\sqrt{\w}(1+\sqrt{\w})^2}\right]}}
\rightarrow  .60 \Phi_2\alpp^{-.366}\\
\Phi_3&=& \Phi_2 a^{-2\w}
\alpp^{{\frac{-w}{1+\w}
\left[1-\frac{2\sqrt{\w}}{(1+\sqrt{\w})^2}\right]}}
\rightarrow  .96 \Phi_2 \alpp^{-.134}\\
\label{eq:ep3pR}
\ep_{3'}&=&\ep_2(\Phi_2/\Phi_3)
\rightarrow  1.01\ep_2 \alpp^{.134}\\
\ep_3&=&\ep_1\sqrt{\w}\sqrt{\rho_3/\rho_1}\rightarrow .62\ep_1 \alpp^{-.232},
\ea
where $\alpp\equiv \rho_1/\rho_2$ and
$a\equiv [(1+\w)/(2\sqrt{\w})]^{1/(1+\sqrt{\w})^2}$, and
where the arrow evaluates the expression when $\w=1/3$.  We note that in
general
$\ep_3\neq\ep_{3'}$.  This is because energy is dissipated over the shock,
increasing
the specific energy for shells behind the shock.
Suppose we take our initial state to be an isentrope so that
$\ep_1/\ep_2=(\rho_1/\rho_2)^{.25}$.
Then, $\ep_3/\ep_{3'}=.6(\rho_2/\rho_1)^{.116}>1$.  Thus the specific energy
and number density
will be discontinuous across the contact discontinuity, and will diverge
as $\rho_1\rightarrow 0$.  In order that this solution be consistent,
$\rho_1/\rho_2$ cannot be too small from \eqr{eq:ep3pR} because
we assumed that $\ep_{3'}/c^2\gg 1$.
In addition, since the temperature in region in region 3 is
$T_3=\w\mu\ep_3$ which we require to be less than the Planck temperature,
the energy density inside the void cannot be too small.
However, we see no reason to restrict $\alpp$ for more general solutions.

The above solution
approximates that of an uncompensated superhorizon-sized
void in a radiation-dominated universe.
We can calculate the approximate
energy density in the ``void'' at times much less than a Hubble time
before collapse.
This solution gives the theoretical maximum limit for the shock tube strength.
Of course, spherical effects will cause the shock strength to increase.
However, our numerical simulations suggest that this increase is
proportional to the initial shock strength.
The ratio of the energy density behind
the shock to that in region 2 is
\be
\label{eq:rho32rel}
\frac{\rho_3}{\rho_2}=\left(\frac{1}{4\beta^2}~\alpp
\right)^{1/(2\beta+1)}
\rightarrow 1.2\alpp^{.536},
\ee
where the arrow evaluates the expression for $\w=1/3$.
In addition, the collapse time is
$\Delta t_{\rm collapse}\simeq \RW/\Phi_1\simeq 2\RW\alpp^{.37}$.
Thus as the ``void'' empties ($\rho_1/\rho_2\rightarrow 0$), the
collapse time decreases rapidly and
the fraction filled before the ``void'' collides at the origin becomes
very small, unless spherical effects are extraordinarily important.

\centerline{\bf{2.  Nonrelativistic, Thin-Walled Uncompensated Void}}

Now we calculate the solution for a shock tube
when region 1 is nonrelativistic: $\rho_1=n_1c^2$ and
$\ep_1/c^2\ll 1$.
We assume that the fluid behind the shock and in the rarefaction wave
are relativistic, however.
We use Eqs.({\ref{eq:UbNR}})-({\ref{eq:shspeedNR}}) with
$\ep_3/c^2\gg 1$, as well as the same rarefaction and contact
discontinuity solutions as was used for the relativistic case in
the previous section.
In particular, we use \eqr{eq:Usim} with $\rho=\rho_3$ and $\rho_0=\rho_2$
in order to obtain $U_{3'}$.
Since $\rho_3=n_3\ep_3=((\w+1)/\w)~\rho_1(\ep_3/c^2)^2$
and $U_3=-\ep_3/c$, $U_3=U_{3'}$ becomes
\be
2\frac{\ep_3}{c^2}\left[\frac{\w+1}{\w}\frac{\rho_1}{\rho_2}
\left(\frac{\ep_3}{c^2}\right)^2\right]^\beta
=
1-\left[\frac{\w+1}{\w}\frac{\rho_1}{\rho_2}
\left(\frac{\ep_3}{c^2}\right)^2\right]^{2\beta}.
\ee
This equation can be solved iteratively to find $\ep_3/c^2$, from which
point the entire solution can be obtained.
We assume now that $\rho_3/\rho_2\ll 1$.
In this limit, we obtain the following solution:
\ba
\label{eq:NRsoltn}
\ep_3/c^2&=&\left\{2 [~\alpp~(\w+1)/\w]^{\beta}
\right\}^{-1/(1+2\beta)}
\rightarrow .50\alpp^{-.232}\\
\label{eq:rho3}
\rho_3&=&[(\w+1)/\w]~\rho_1~\ep_3^2/c^4
\rightarrow \rho_1 \alpp^{-.464}\\
U_3/c&=&-\ep_3/c^2\\
\label{eq:Phi1NR}
\Phi_1&=&2\w  \Phi_2 \left(\frac{\w+1}{4\w}\right)^{\sqrt{\w}/(1+\sqrt{\w})^2}
\alpp ^{{\frac{-w}{1+\w}
\left[1+\frac{1-\w}{\sqrt{\w}(1+\sqrt{\w})^2}\right]}} \rightarrow  (2/3)\Phi_2
\alpp^{-.366}\\
\Phi_3&=&\Phi_1/\left[(\w+1)\ep_3/c^2\right]
\rightarrow  \Phi_2\alpp^{-.134}\\
\label{eq:ep3p}
\ep_{3'}&=&\ep_2(\Phi_2/\Phi_3)
\rightarrow  1.0\ep_2 \alpp^{.134},
\ea
where $\alpp\equiv \rho_1/\rho_2$.
This solution depends on $\alpp$ in the same way
as it does in the relativistic case.
The prefactors however, are somewhat different.
Using \eqr{eq:rho1lim}, ({\ref{eq:rho3}}), ({\ref{eq:Phi1NR}})
and ({\ref{eq:tcoll}}), we find a rough lower bound for the collapse
time
\be
\Delta t_{\rm collapse}>.1c^{-1}\RW(\mu/M_{PL})^{.633}.
\ee
where $\mu$ is the particle mass.
(In general, because $T_{3'}<M_{PL}$,
we can estimate the lower bound on the collapse time
using \eqr{eq:PhiNR1} to be
$\Delta t_{\rm collapse}> c^{-1}\RW~\mu/M_{PL}$).
We can also trace out the boundaries between regions, and the radii of all
comoving shells.  $X_a$ and $X_c$ are still given by
Eqs.({\ref{eq:Xa}}) and ({\ref{eq:Xc}}), respectively.  However, now
\ba
X_b&=&\XW+(t-\ti)\Phi_3 U_3(1-\sqrt{\w})\\
X_d&=&\XW+\Phi_3 U_{\rm sh}(t-\ti)=\XW- c(t-\ti)\Phi_1.
\ea
The rarefaction wave moves out at the speed of sound, and the shock
moves in at roughly the speed of light.

\centerline{{\bf C. Comparison of the Collapse with the}}
\centerline{{\bf Shock Tube Solutions---Numerical Results}}

In this section, we find the numerical solutions to the special relativistic
equations of motion
with spherical symmetry and compare them to the
slab-symmetric shock tube solutions derived in the last section.
For {\it Figures 9-12} and {\it Table 5}, we set
$C=.3$, $c=1$, $\w=1/3$, $\ti=1$, $G_N=0$,
$U(\ti,R)=0$ and $4\pi\rho_{\rm out}(\ti)=3/8$.

In {\it Figure 9} we show the collapse of a relativistic,
uncompensated void
with $\RW=20$, $\rw=.005$, $\DR=.0025$, $\alpha =10^{-2}$,
$\ep_{\rm out}(\ti)/c^2=10^6$ and $k^2=2.5$.
The triangles show the pressure and specific energy
as a function of $R$ at $\ti$ and $t=1.5$.
As usual, the
void collapses via a shock moving toward the origin at the
speed of light.
The shock tube solution is plotted as dashed lines.
The agreement is excellent, because
the shock has not traveled very far
{\it and} $|\log_{10}\alp|\rw/\RW\ll 1$.

Due to time dilation in the void,
light travels a farther distance inside the void than outside it in the same
amount of coordinate time $t-\ti$.  Therefore, the amount of time taken to
form the shock is very important for relativistic fluids.
If the wall is too thick, the shock will take longer to form.
In addition, the shock strength will be smaller,
decreasing the void's relative potential and therefore
the time dilation effect.
This results in the solution not equaling the shock tube solution at any time.
In {\it Table 5}, we show the location of the shock, $R_{\rm sh}$, at
$t=1.5$ for voids
with the same initial conditions as that in {\it Figure 9} but for
varying wall thicknesses and $\DR=\rw/2$.
In addition, the value of $\Phi_{\rm in}$ is shown.
Note that $\Phi_{\rm in}(\ti)=3.16$ for all voids.
It is clear that for this value of
$\alpha$, the wall must be thinner than
$2\rw/\RW\la 2\times 10^{-3}$ in order that the solution
approximate the shock tube well.

In {\it Figure 10} we plot $p$, $\ep$ and $U$ as a function of time
for the void in {\it Figure 9} but at $t=5$.
Note the large distortion in pressure, specific energy and velocity behind the
shock.
This is due to spherical geometrical effects; fluid behind the shock is forced
to occupy
a smaller volume, thereby increasing the number density.
This strengthens the shock, which in turn dissipates more energy so that
the specific energy, energy density and velocity increase.
In addition, it is seen that the shock is far ahead of the
shock-tube shock.  This is a consequence
of the increased energy density behind the shock, $\rho_b$.
{}From Eqs.({\ref{eq:phibr}}) and ({\ref{eq:Phirel}}), the value
of $\Phi$ inside the void is
\be
\Phi_{\rm in}=\frac{1}{\sqrt{3}}
\left(\frac{\rho_{\rm out}}{{\rho_{\rm in}}^2}\right)^{1/4}
\rho_b^{1/4}.
\ee
Therefore, as the energy density behind the shock increases,
the shock moves further per unit coordinate time because
$dR_{\rm sh}=c\Phi dt$.

In {\it Figure 11(a)}, we show the shock tube solution for $\alp=10^{-3}$
and $\XW=10$.
The dotted lines are the position of
$X_a,~X_B,~X_c$ and $X_d$, and the solid lines
are the locations of comoving observers with
$X(\ti,x)=11.0$, $10.1$, $10.0$, $8.0$, $4.0$ and $2.0$.
Note that the shock moves very far per coordinate time interval,
due to the time dilation
effect.
Note also that after entering regions $3$ or $3'$, all shells move with
the same velocity.
In {\it Figure 11(b)}, we show the comoving shell positions again as solid
lines.
In addition, the dashed lines are the positions of comoving shells with the
same
initial locations, $R(\ti,r)=X(\ti,x)$, in an evolving {\it void}.
For this void,
$\RW=10$, $\rw=.1$, $\DR=.025$, $\alpha =10^{-3}$,
$\ep_{\rm out}(\ti)/c^2=10^6$ and $k^2=1.7$.
Although the void wall is not very thin,
the solution approximates the shock tube solution fairly well
in region $3$.  However, it does not do as well
in regions $3'$ and $4$, due to effects stemming from the thick wall.
Note that the numerical shock collides at the origin at $t=2.2$, which
is why the velocity of the comoving shell with $R(\ti,r)=2.01$
changes after this time.

In {\it Figure 12}, we plot $p$, $\ep$ $n$ and $U$ for the collapse of a
nonrelativistic void in a relativistic background:
$\ep_{\rm in}(\ti)/c^2=.38$, and $\ep_{\rm out}(\ti)/c^2=5.0$.
In addition, $\RW=20$, $\rw=.0025$, $\DR=\rw/4$, $\alpha =10^{-4}$
and $k^2=3.0$.
The triangles are the numerical solution at $t=1.06$.\footnote{Note that
the numerical simulation covers the range $R(\ti,r)\in [19.2,20.1]$.
In addition, $\rw$ is chosen to be small enough that the wall acts
like an initial discontinuity.}
The general features are similar to the
relativistic void solution: the void collapses at the speed of light as
the wall fluid shocks inward.
In addition,
a contact discontinuity (where $p$ and $U$ are continuous,
while $\ep$ and $n$ are discontinuous)
follows behind the shock.  Finally, a rarefaction wave
moves into the region outside the void.
Note that the shock and contact discontinuity are located
at $R=19.33$ and $19.47$, respectively, while the rarefaction
wave is located between $R=19.7$ and $20.04$.

The dashed line is a plot of the nonrelativistic/relativistic shock tube
solution
derived in section B2.
It is clear that this solution agrees well with the numerical solution
in the
outer part of the rarefaction wave (large $R$).
In addition, $p$, $\ep$ $n$ and $U$  are predicted fairly well
behind the shock.
(For example, $\rho_3/\rho_1=61$ and $\ep_3=4.2$ from the numerical solution,
while $\rho_3/\rho_1=72$ and $\ep/c^2=4.2$
from Eqs.({\ref{eq:rho3}}) and ({\ref{eq:NRsoltn}})).
However, because the fluid in regions $3'$ and part of $4$ are barely
relativistic
($\ep/c^2\simeq 1$), the assumptions used to derive the
shock tube solution are not satisfied in this case.
Therefore, it is no surprise that the shock tube solution
does not match the numerical solution very well.
At $t=1.06$, $\Phi_{\rm in}=11.2$ from the numerical solution,
while $\Phi_1=19.4$
from \eqr{eq:Phi1NR}.
Note that $\Phi_{\rm in}$ increases from its initial
value $\Phi_{\rm in}(\ti)\simeq \ep_{\rm out}/c^2 = 5.1$
because energy is dissipated over the shock.
This results in a smaller collapse time.

Thus, the relativistic shock tube solution
approximates the numerical solution of a thin-walled
void very well for a significant amount of time during collapse.
The nonrelativistic/relativistic
shock tube solution does not do as well because
the fluid between the contact discontinuity and the rarefaction wave
tends to be barely relativistic or slightly nonrelativistic.
However, it is qualitatively (and to some degree quantitatively) correct.

\vspace{.2in}

\centerline{\bf{VI. Discussion}}
\setcounter{section}{6}
\setcounter{equation}{0}

In this paper, we studied in detail
the non-linear collapse of a superhorizon-sized void embedded
in a background radiation-dominated
FRW universe.  We find in general that a relativistic
or nonrelativistic void collapses
via an inbound shock at the speed of light.
This occurs because the pressure and velocity gradients in the lower part
of the void wall are enormous and therefore accelerate fluid into the void.
Because of a time dilation effect, this occurs very quickly, and
in particular can occur in less than an outside Hubble time
if the relative energy density inside a void is
small enough.  This is true even
when the outward velocity of the wall is enormous.
A large wall velocity does drive some of the fluid in the
wall out behind a shock, so that less fluid ends up
in the void.
This causes the strength of the inbound shock to decrease, and
lessens the relative potential of the void.  Thus
the collapse time increases somewhat.
However, the void can still collapse in less than an outside Hubble time.
This is important, since it emphasizes the point that seemingly nothing
stops the void from quickly collapsing in on itself.

As the void wall gets thinner, the shock strength increases and therefore the
collapse time decreases.  This behavior does not extrapolate to
infinitely thin walls, however.  When the wall becomes sufficiently thin
so that it acts like an initial discontinuity,
a maximum value for the shock strength
(and therefore a minimum value for the collapse time) is reached.
Up to the point when the void has partially collapsed, the solution
can be approximated by the shock tube solution.
This holds for an
uncompensated, superhorizon-sized void which collapses in much less
than an outside Hubble time and for which the fluid in the void is
relativistic or nonrelativistic.
If the fluid in the void is nonrelativistic, then
then the collapse takes longer to occur for the same
value of $\rhoi/\rhoo$.

If the wall velocity is small and the void collapses in less than
an outside Hubble time, then the solution at the collapse time
is virtually the same as that when gravitational effects are neglected.
This is because fluid in the lower part of the wall is accelerated into
the void by pressure and velocity gradient forces, not gravitational
forces, and the fluid configuration in the peak area is virtually unchanged.

When the wall velocity is large however, it is not obvious a priori
whether or not gravitational forces are important in moving the
peak area fluid out of the void wall.
In this case,
fluid in the peak area of the wall moves out in
less than an outside Hubble time.  In addition, the time scale
for this to occur is proportional
to $\rw^{3/4}/\GW$, where $\rw$ is the wall thickness,
and $\GW$ is approximately the ``gamma-factor'' or ``velocity'' of the wall
when $\GW > c^{-1}\RW/\Hm$.
(If $G_N=0$, $\GW$ is the relativistic gamma-factor of the fluid,
and equals the ``velocity'' when $\GW\gg 1$).
Therefore, if $\GW$ is large enough,
fluid in the peak area will
shock outward during the collapse, distoring the shape of the wall in
the process.
It is found however, that in general this does not change the collapse
time by very much, given a fixed value of
$\rhoi/\rhoo$.  In addition,
it is found that
gravity is unimportant in shaping the distortion
of the fluid configuration in the peak.  Therefore,
gravitational contributions to the solution at the collapse time
are negligible.  This means that only non-gravitational forces
are driving
the fluid into and out of the void wall, even though the void is
superhorizon-sized.
(Although it is clear that pressure and velocity gradient forces must
contribute to
driving the fluid out of the peak
area, one might envision gravitational forces changing the configuration
by substantially redshifting the wall pressure, as it does when the collapse
occurs in greater than an outside Hubble time.
Apparently, gravitational forces do not act quickly enough to do this.)
This sounds surprising, especially since superhorizon-sized voids are larger
than the Hubble radius and therefore must be influenced heavily
by gravitational forces.  It is
a question of time scales, however.
Because time is dilated in the void, if the collapse occurs in much
less than an outside Hubble time, gravity has not had a chance to
substantially change the solution (pressure distribution, shell locations, etc)
at the collapse time in this case.
Gravitational forces should be extremely important
{\it after} the void collapses, of course, which
was not studied in this paper.
Then, they will try to pull fluid lumps together, while
fluid forces will try to pull fluid lumps apart.

There are many questions still left to answer.
What happens to a superhorizon-sized void {\it after} it collapses?
Now that we have a code which implements numerical regularization
to prevent instabilities near the origin, it will be
possible to explore this late stage numerically.
In SV, we found that after a special relativistic void collapses,
the pressure near the origin is
much greater than the pressure outside the void, creating
an overdensity.  Because
this lump drops off with distance from the origin, this
large pressure gradient accelerates fluid away from the origin.
What happens if gravity is added to this problem?
It appears likely that
a black hole might form if the wall velocity is not
very large.  If it is very large, enough
fluid in the peak area might
be pushed out of the void initially so that this is prevented.
In this case, a mass-energy deficit will occur in the void initially.
If this mass-energy continues to be pushed away from the void, thermalization
and homogenization of this void might take a long time to occur.
However, as this expelled
fluid slows down, its mass-energy $M$ will decrease.
In addition, at the collapse time the velocity in the former void region
is negative, so its mass-energy will increase.
Thus, it will be important
to analyze the long-term solution to see if there is a mass deficit
created where the void was located.  In addition,
density waves will be created, and must also be followed for their long-term
behavior---do they become growing density perturbations, or remain
outgoing local waves?
A companion paper will explore these questions\re{Vasha1}.

\vspace{18pt}

\centerline{\bf ACKNOWLEDGMENTS}

The author was supported by the President's Postdoctoral Fellowship Program
at the University of California, and NSF Grant AST-9120005.
She would like to thank A. Liddle, P. Steinhardt and Peter Anninos for
stimulating and useful discussions.

%%%%%%%%%%%%%%%%%%%%%%%%%%%%%%%%%%%%%%%%%%%%%
%%%%%%%%%%%%%%%%%%%%%%%%%%%%%%%%%%%%%%%%%%%%%
\frenchspacing
\def\prl#1#2#3{Phys. Rev. Lett. {\bf #1}, #2 (#3)}
\def\prd#1#2#3{Phys. Rev. D {\bf #1}, #2 (#3)}
\def\plb#1#2#3{Phys. Lett. {\bf #1B}, #2 (#3)}
\def\npb#1#2#3{Nucl. Phys. {\bf B#1}, #2 (#3)}
\def\apj#1#2#3{Astrophys. J. {\bf #1}, #2 (#3)}
\def\apjl#1#2#3{Astrophys. J. Lett. {\bf #1}, #2 (#3)}
%%%%%%%%%%%%%%%%%%%%%%%%%%%%%%%%%%%%%%%%%%%%%
%%%%%%%%%%%%%%%%%%%%%%%%%%%%%%%%%%%%%%%%%%%%%
\begin{picture}(400,50)(0,0)
\put (50,0){\line(350,0){300}}
\end{picture}

\vspace{0.25in}

\def\labelenumi{[\theenumi]}

\begin{enumerate}

\item\label{LASTEIN} D. La and P. J. Steinhardt, {\em Phys. Rev. Lett.}
	{\bf 62}, 376 (1989); {\em Phys. Lett.} {\bf 220B}, 375 (1989);
	F.S. Acceta and J.J. Trestor, Phys. Rev. D {\bf 39}, 2854 (1989);
	A. Linde, Phys. Rev. D {\bf 49}, 748 (1994);
	R.Holman et al, Phys. Rev. D {\bf 43}, 3833-3845 (1991);
	F.C.Adams and K. Freese, Phys. Rev D {\bf 43}, 353 (1991);
	J. Garcia-Bellido and M. Quiros, Phys. Lett. {\bf B 243}, 45 (1990);
	E.J. Copeland et al, Phys. Rev. D {\bf 49}, 6410 (1994) and
	references therein.

\item\label{BigBubble} E. J. Weinberg, {\em Phys. Rev. D} {\bf 40},
	3950 (1989); D. La, P. J. Steinhardt, and E.W. Bertschinger,
         {\em Phys. Lett.} {\bf 231B}, 231 (1989);
	M.S. Turner, E.J. Weinberg, L.M. Widrow, {\em Phys. Rev. D} {\bf 46},
	2384 (1992); A.R. Liddle and D. Wands, {\em Mon. Not. Roy. Aston. Soc.},
	{\bf 253}, 637, (1991).

\item\label{Vasha} S.L. Vadas, {\em Phys. Rev. D} {\bf 48}, 4562 (1993);
``The Evolution of Superhorizon-sized Voids in the
Early Universe,'' S.L. Vadas, Numerical Simulations in Astrophysics conference
proceedings, July, 1993, (in press).

\item\label{Vasha1} S.L. Vadas, ``Collapse of a Superhorizon-sized Void in the
Early
Universe, II'' (unpublished).

\item\label{MisSharp} C.W. Misner and D.H. Sharp, {\em Phys. Rev }, {\bf 136},
B571, (1964);
	W.C. Hernandez, Jr. and C.W. Misner, {\em Astro. J.} {\bf 143}, 452, (1966);
	M.M. May and R.H. White {\em Phys. Rev}, {\bf 141}, 1232, (1966).
\item\label{MayWhite} M.M. May and R.H. White, {\em Meth. Computat. Phys},
	{\bf 7}, 219 (1967).
\item\label{VonNRich} J. VonNeumann and R.D. Richmyer,
	{\em Jour. of Applied Physics},	{\bf 21}, 232, (1950).
\item\label{Evans} C.R. Evans, ``An Approach for Calculating Axisymmetric
	Gravitational Collapse'', from Dynamical Spacetimes in Numerical
	Relativity, edited by J. Centrella, Cambridge Univ. Press, 1986;
	C.R. Evans, L.L. Smarr and J.R. Wilson, ``Numerical Relativistic
	Gravitational Collapse with Spatial Time Slices'',
	from Astrophysical Radiation Hydrodynamics'', Edited by K-H Winkler
	and M. Norman, Reidel Publ. Co., 1986.

\item\label{FOBub} S.L.Vadas (unpublished),
	``Thermalization of a Large Bubble's Wall after Inflation,''
	S.L. Vadas, S. Davidson and J. Cohn (unpublished).

\item\label{Stoneetal} J.M. Stone et. al., {\em Astrophys. J.},
	{\bf 388}, 415 (1992).

\item\label{Sedov} L.I.Sedov, {\em Similarity and Dimensional Methods in
	Mechanics}, Academic Press, New York, 1959.

\item\label{CahTaub} M.E. Cahill and A.H. Taub, {\em Commun Math. Phys},
	{\bf 21}, 1 (1971).

\item\label{ShockTube} J. Centrella and J.R. Wilson,
	{\em Astro. J. Suppl.}, {\bf54}, 229 (1984);
	K.W. Thompson, {\em J. Fluid Mech}, {\bf 171}, 365 (1986).

\item\label{Synge} J.L. Synge, {\em Relativity: The General Theory},
	North-Holland Publ., Amsterdam (1960).

\item\label{CourFrei} R. Courant and K.O. Friedrichs, {\em Supersonic Flow and
	Shock Waves}, Springer-Verlag, New York, (1985).

\end{enumerate}

\nonfrenchspacing
\centerline{\bf Figure Captions}

{\bf Fig.~1:} Relative error in energy density,
$(\rho-\rho_{\rm hom})/\rho_{\rm hom}$, where
$\rho_{\rm hom}$ is the FRW solution, versus $R\sqrt{\ti/t}$
for 3 different values of $\f$.
The improved code does much better than that used in SV.

\vspace{10pt}

{\bf Fig.~2:} The relative difference in pressure as a function of initial
radii for the collapse of GR and SR
voids with nearly identical initial conditions.  $\GW=1$ for both.
The closed triangles, open
squares, and dashed line show the results
for $\alp=10^{-4}$,
$\alp=10^{-7}$ and $\alp=10^{-10}$.
As the collapse time becomes much less than an outside Hubble radius
($\alp\la 10^{-10}$),
the GR and SR solutions are very similar.

\vspace{10pt}

{\bf Fig.~3:} Same as {\it Figure 2}, but for $\GW=6$.

\vspace{10pt}

{\bf Fig.~4:} The evolution of six comoving observers as a function
of time for the voids from {\it Figure 3} with $\alp=10^{-4}$(top)
and $\alp=10^{-10}$(bottom):
$R(\ti)=30,~35,~40,~45,~48,~50$, $55$ and $60$.
The solid (dashed) lines are the GR (SR) solutions.
These solutions are very similar for $\alp=10^{-10}$ because
the collapse occurs in much less than an outside Hubble time.

\vspace{10pt}

{\bf Fig.~5:} $4\pi\rho(t)$ and $\Gamma(t)$ for 3 comoving
shells in the peak area, $R(\ti)=50,~49$ and $48$ (top to bottom)
for the voids from {\it Figure 3}
with $\alp=10^{-4}$.
The GR (SR) void results are shown as solid (dashed) lines.  Note that
the peak energy density redshifts away quite quickly for the GR,
but not for the SR, void.

\vspace{10pt}

{\bf Fig.~6:} $4\pi p$ and $\Gamma$ as a function of radius for a
superhorizon-sized void
25 times the Hubble radius and with $\alp=10^{-10}$.
The initial distributions are shown as solid lines.
The numerical GR (SR) results are plotted as dotted (dashed) lines.
Note that the peak has moved
outward substantially at $\Delta t/\Hm=.03$ for both voids, but
that the solutions are very similar anyway.

\vspace{10pt}

{\bf Fig.~7:} Same as {\it Figure 6} but for voids with relative energy density
inside the void $1000$ times smaller.  Note that the GR
and SR solutions are nearly identical at $\Delta t/\Hm=.005$, even
though the peak area has shocked outward substantially.

\vspace{10pt}

{\bf Fig.~8:} Schematic diagram of the shock tube showing the different
regions.  The shock
is located at $X_d$, and is moving into region 1, and the outer edge
of the rarefaction wave is located at $X_a$, and is moving into region 2.

\vspace{10pt}

{\bf Fig.~9:} Pressure and specific energy as a function of radius
for a thin-walled SR void.  The initial distributions are the solid
lines, the numerical solution is the triangles, and the shock tube solution is
the dashed lines.  Note that the numerical solution is well
approximated by the shock tube solution.

\vspace{10pt}

{\bf Fig.~10:} Same void as in {\it Figure 9}, but at a much later time.
The numerical solution is given by the solid lines in this figure.
The pressure, specific energy and velocity
behind the shock is much larger than the shock tube solution.  This is
due to spherical geometrical effects.  The upper part of the rarefaction
wave however, is unaffected by these effects.

\vspace{10pt}

{\bf Fig.~11:}{\bf (a)} Location versus time for the shock tube solution with
$\alp=10^{-3}$ and $\RW=10$.
The position of the $4$ boundary locations are the dotted lines, and the solid
lines are the locations of comoving points with
$X(\ti)=2,~4,~8,~10,~10.1$ and $11$.
{\bf (b)}  Shock tube solution (solid lines) and numerical solution
(dashed lines) for the same comoving points from {\it Figure 11(a)}.
The two solutions coincide well in region 3, but not as well in the
wall area ($X(\ti)=R(\ti)=10,~10.1$) because the thick wall
evolves differently than that given by
the simple similarity solution.

\vspace{10pt}

{\bf Fig.~12:} Solution for an uncompensated, nonrelativistic void
in a relativistic background (triangles).  The dashed lines
show the nonrelativistic/relativistic shock tube solution.

\vspace{10pt}

{\bf Fig.~13:} Percent change in $\Phi$, $\ep$, $\Gamma$ and $n$
as a function of shock strength for a general relativistic fluid.
Note that $\ep$ is underestimated,
while the other quantities are overestimated, at the shock.

%\vspace{1in}
\newpage
\vspace{.2in}

\centerline{\bf{Appendix A:  Contact Discontinuity and Shock Jump Conditions}}
\@addtoreset{equation}{section}
\def\ksection{\arabic{section}}
\def\theequation{A.\arabic{equation}}
\setcounter{equation}{0}

In this section, we briefly re-derive the jump conditions for general
relativistic
shocks.  This closely follows Ref.\re{MayWhite}.
In addition, we extend the derivation to include the conditions
satisfied over a contact discontinuity.
Let the variables $a$ and $b$ represent labels for comoving observers
ahead and
behind a discontinuity, respectively.\footnote{Here, ``discontinuity''
refers to either a contact discontinuity or a shock.}
If each observer measures the invariant
interval separating the same two events on the world surface of
this discontinuity,
using \eqr{eq:metric}, we obtain
\be
\label{eq:shmet}
[ds^2]=[-c^2\Phi^2 dt^2+\Lambda^2dr^2+R^2d\Omega^2]=0,
\ee
where $[G]\equiv G_a-G_b$.  Because the discontinuity is
radial, we can choose the two events to have
$dr=dt=0$ but $d\Omega\neq 0$.
Therefore, $R$ is continuous across the discontinuity: $[R]=0$.
Now consider two events with $dr\neq 0$ and $dt\neq 0$.  Then from
\eqr{eq:shmet},
\be
\label{eq:phi2}
[c^2\Phi^2-M_{\rm sh}^2/n^2]=0,
\ee
where $r_{\rm sh}$ is the position of the discontinuity in comoving
coordinates,
$dr_{\rm sh}/dt$ is the discontinuity ``speed'', and
$M_{\rm sh}\equiv {f ~(dr_{\rm sh}/dt)}/({4\pi R^2})$ from
Eqs.({\ref{eq:defnf}}) and ({\ref{eq:Gamlam}}).
If the discontinuity is a contact discontinuity, then $r_{\rm sh}=const$,
or $dr_{\rm sh}/dt=0$.

It can be shown that the conditions satisfied over the discontinuity
for the Schwarzschild metric (but {\it not} for the comoving metric)
are\re{Synge},\re{MayWhite}
\be
\label{eq:cont}
[{T_\nu}^\mu\partial g/\partial x^\mu]=0,
\ee
where $g$ is the equation for the world surface of the discontinuity.
Therefore we first find these expressions
in the Schwarzschild frame and then transform back
to the comoving frame.
The Schwarzschild metric is
\be
ds^2=-c^2 A^2 dT^2+B^2dR^2+R^2d\Omega^2,
\ee
where $R$ is now the ``Eulerian'' coordinate radius.  Since $R$ is
a coordinate, $[R]=0$ over the discontinuity, and using the same argument as
above for the two observers,
\be
\label{eq:new}
[A^2-B^2S^2]=0,
\ee
where $S$ is the discontinuity ``speed''
in this frame: $cS\equiv dR_{\rm sh}/dT$.  For the Schwarzschild metric, the
solution for the metrics functions are well known: $B^{-2}=1-2MG/(Rc^2)=A^{2}$
where $M(R,T)=4\pi c^{-2}\int_0^R\rho R^2dR$.
As long as $\rho$
is not infinite in the discontinuity,
the mass is continuous across it, $[M]=0$.
Then $[B]=[A]=0$, and from \eqr{eq:new},
$[S]=0$.

We now derive these conditions in the Schwarzschild frame.
If a discontinuity is located at position $R_{\rm sh}(T)$ at time $T$, the
equation for its
world surface is
$g=R_{\rm sh}(T)-R=0$.
In addition, the perfect fluid stress-energy tensor in the Schwarzschild
frame\footnote{We
denote the metric and $4-$velocity in the Schwarzschild frame by primes.} is
\be
{T'_\mu}^\nu=c^{-2}(\rho+p)g'_{\mu\lambda}u'^\nu u'^\lambda+pg'_{\mu\lambda}
g'^{\nu\lambda}=
n{\rm W}g'_{\mu\lambda}u'^\nu u'^\lambda+pg'_{\mu\lambda}g'^{\nu\lambda},
\ee
and the conservation of mass equation is
$[nu'^\nu \partial g/\partial x^\nu]=0 $\re{Synge},\re{MayWhite}.
Using this and \eqr{eq:cont}, the junction conditions become
\ba
\label{eq:eqen}
c^2\left[(-A^2(u'^T)^2n{\rm W}+p)S+n{\rm W}A^2u'^Tu'^R\right]&=&0\\
\label{eq:eqmom}
\left[Sn{\rm W}B^2u'^Tu'^R-(n{\rm W}B^2(u'^R)^2+p)\right]&=&0\\
\label{eq:eqma}
\left[nSu'^T-nu'^R\right]&=&0.
\ea

We would like to express these equations in terms of
the comoving metric functions.  First,
we can rewrite the speed in the Schwarzschild frame
as $cS=(R_t+R_r \rs)/(T_t+T_r\rs)$.  In addition, we can
relate the metric functions from the comoving frame to those in the
Schwarzschild frame by
$g'^{\mu\nu}=({\partial x'^\mu}/{\partial x^\sigma})~
({\partial x'^\nu}/{\partial x^\lambda})~g^{\sigma\lambda}$.
We then obtain $(cA)^{-2}=(c\Phi)^{-2}T_t^2-\Lambda^{-2}T_r^2$,
 $B^{-2}=-(c\Phi)^{-2}R_t^2+\Lambda^{-2}R_r^2$ and $(c\Phi)^{-2}T_tR_t=
\Lambda^{-2}T_rR_r$.
In addition, the ``mass-energy'' $M(r,\t)$ is
continuous across the discontinuity,
since $[M]=4\pi c^{-2}[\rho R^2]dR~^{~~~\Rightarrow}_{dR\rightarrow 0} ~0$.
With this and \eqr{eq:defGamma}, we find that $[\Gamma^2-U^2/c^2]=0$.

We can use these relations to calculate the 4-velocity as
measured in the Schwarzschild frame.
Since the 4-velocity in the comoving frame is $u^\mu=(-c\Phi^{-1},0,0,0)$,
and the velocity transforms as
$u'^\lambda=(\partial x'^\lambda/\partial x^\sigma)~ u^\sigma$,
$u'^\lambda=-\Phi^{-1}(c\dot{T},\dot{R},0,0)$,
where $\dot{T}=\partial T/\partial t$ and $\dot{R}=\partial R/\partial t$.
Using the normalization condition
$u'^\lambda u'_\lambda=-c^2=-c^2A^2(u'^T)^2+B^2(u'^R)^2$ along with
$U\equiv \dot{R}/\Phi$, we can
rewrite the $4$-velocity as
$u'^\lambda=-(A^{-1}\sqrt{B^2U^2+c^2},U,0,0 )$.
In the special relativistic limit ($G_N=0$), $A=B=1$.
The $4-$velocity of a fluid particle is then
\be
\label{eq:Gaminterp}
u'^\lambda = -((U^2+c^2)^{1/2},U,0,0)= -(c\Gamma,\Gamma {v},0,0),
\ee
where we have defined ${v}\equiv U/\Gamma$ so that
$\Gamma=1/\sqrt{1-(v/c)^2}$.
In the special relativistic limit, then, ${v}$ is the fluid
particle's radial velocity, $\Gamma$ is usual gamma-factor (i.e. energy per
particle mass),
and $U$ is the particle momentum per particle mass.

Across a contact discontinuity, which is
a discontinuity with $\dot{r}_{\rm sh}=0$,
Eqs.({\ref{eq:phi2}}), ({\ref{eq:eqen}})-({\ref{eq:eqmom}})
and $[S]=0$,
\ba
\label{eq:Ph}
[\Phi]&=&0\\
\label{eq:pUG}
[pU/\Gamma]&=&0\\
\label{eq:p}
[p]&=&0\\
\label{eq:UG}
[U/\Gamma]&=&0
\ea
respectively.\footnote{\eqr{eq:eqma} gives $[0]=0$.}
Thus the pressure is continuous across a contact discontinuity,
as it is in the nonrelativistic case.
Using \eqr{eq:defGamma} and ({\ref{eq:UG}}), we find
that $(U_a^2-U_b^2)(1-2G_NM/(Rc^2))=0$.  If $2G_NM/(Rc^2)\neq 1$,
$[U]=0$, or the tangential velocity is continuous across the contact
discontinuity,
as it is in the nonrelativistic case.
If $2G_NM/(Rc^2)=1$, however, the entire formalism breaks down because
then $B=\infty$.  This occurs at the Schwarzschild radius.
Thus, across a contact discontinuity,
\be
\label{eq:contact}
[\Phi]=0,~~~ [p]=0,~~~[U]=0.
\ee

We now find the jump conditions at a shock where $\dot{r}_{\rm sh}\neq 0$.
We rewrite Eqs.({\ref{eq:eqen}})-({\ref{eq:eqma}})
and $[S]=0$ in terms of the
comoving metric functions using the fact that
$T_r/T_t=c^{-2}U\Lambda/(\Gamma\Phi)$ and $[\rs]=0$. We obtain
\ba
\label{eq:ma}
[U M_{\rm sh}/(nc^2)+\Phi\Gamma]&=&0\\
\label{eq:en}
[c^2M_{\rm sh}\Gamma(1+\ep/c^2)-pU\Phi]&=&0\\
\label{eq:mom}
[M_{\rm sh} U(1+\ep/c^2)-p\Gamma\Phi]&=&0\\
\label{eq:shvel}
[M_{\rm sh}\Gamma/n+\Phi U]&=&0.
\ea
\eqr{eq:phi2} and Eqs.({\ref{eq:ma}})-({\ref{eq:shvel}})
make up the required shock conditions.
One of them however, is redundant.
It can be shown that in the nonrelativistic limit,
the above conditions reduce to the
Lagrangian shock jump conditions given in Ref.\re{CourFrei}.

We are interested in the case where the fluid in front of the shock
is either nonrelativistic or relativistic.  First, consider the
nonrelativistic
case, which was considered previously\re{MayWhite}.
We set $G_N=0$ and $\ep_a=p_a=U_a=0$ and take
$p_b=\w\ep_b n_b$.
Using \eqr{eq:phi2} and Eqs.({\ref{eq:ma}})-({\ref{eq:shvel}}),
we find that
\ba
\label{eq:UbNR}
U_b/c&=&\mp\sqrt{\Gamma_b^2-1}\\
\eta&\equiv& \frac{n_b}{n_a}=\left[2+\w+(\w+1)\ep_b/c^2\right]/\w \\
\Phi_b&=&\frac{\Phi_a}{1+(\w+1)\ep_b/c^2}\\
\Gamma_b&=&1+\ep_b/c^2\\
\label{eq:shspeedNR}
R_{\rm sh}'\dot{r}_{\rm sh}&=&\mp c\Phi_a\sqrt{\frac{\ep_b/c^2}
{\ep_b/c^2 +2}}
\left(\frac{2+\w+(\w+1)\ep_b/c^2}{1+(1+\w)\ep_b/c^2}\right).
\ea
Thus, all quantities behind the shock can be expressed in terms of
the specific internal energy behind the shock, $\ep_b$.
Note that when the fluid behind the shock is relativistic,
$\ep_b/c^2\gg 1$, the shock moves at
the speed of light, $R_{\rm sh}'\dot{r}_{\rm sh}=\mp c \Phi_a$.

The energy density ahead of the shock
cannot be zero since $n_a>0$.  If the mass density were zero, then
$\ep_b=\infty$.  But this is impossible since $\w\ep_b=T_b/\mu$,
where $\mu$ is the particle mass,
leading to an infinite temperature behind the shock.  Therefore, we
require $T_b<M_{PL}$ in order that the classical field
equations be valid.
Then, if $\ep_b/c^2\gg 1$,
the energy density in front of the shock is bounded from below:
\be
\label{eq:rho1lim}
\rho_a=n_a c^2 > [\w^3/(\w+1)]~\rho_b (\mu/M_{PL})^2.
\ee
In addition, it can be shown using
Eqs.({\ref{eq:en}}) and ({\ref{eq:mom}}) that setting $n_a=0$ leads
to $U_b^2-\Gamma_b^2=0$,
in direct contradiction with \eqr{eq:UbNR}.

When the fluid in front of the shock is relativistic, we end
up with a completely different set of equations.
We first set $G_N=0$ and $U_a=0$.
In addition, $p=\w\rho$ and $\rho=n\ep$.
Then \eqr{eq:phi2} and Eqns({\ref{eq:ma}})-({\ref{eq:shvel}})
yield
\ba
\label{eq:Ur}
U_b&=&\mp\sqrt{\Gamma_b^2-1}=\mp \frac{\sqrt{\w}}{\w+1}\frac{\rho_b-\rho_a}
{\sqrt{\rho_a\rho_b}}\\
\label{eq:Gr}
\Gamma_b&=&\sqrt{\frac{(\w\rho_a+\rho_b)(\rho_a+\w\rho_b)}
{(w+1)^2\rho_a\rho_b}}\\
\eta&\equiv&\frac{n_b}{n_a}=\frac{(1+\w)\Gamma_b\rho_b}{\w\rho_b+\rho_a}\\
\label{eq:phibr}
\Phi_b&=&\frac{\ep_a}{\ep_b}\Phi_a=\Phi_a\sqrt{
\frac{(\w\rho_a+\rho_b)\rho_a}{(\w\rho_b+\rho_a)\rho_b}}\\
\label{eq:shspeed}
R_{\rm sh}'\dot{r}_{\rm sh}&=&
\mp (1+\w)\frac{\rho_b}{\rho_b-\rho_a} \Phi_b U_b.
\ea
We note from Eqn~({\ref{eq:phibr}}) that $[\Phi w]=0$.

In the strong shock limit, when $\rho_b\gg\rho_a$, the shock moves at
the speed of light: $R_{\rm sh}'\dot{r}_{\rm sh}\simeq \pm c\Phi_a$.
In addition, $n_b/n_a\simeq\sqrt{\rho_b/(\w\rho_a)}$ and
$\ep_b/\ep_a\simeq\sqrt{\w\rho_b/\rho_a}$.  Thus,
$n$ and $\ep$ do not scale as $\rho^{\w/(1+\w)}$ as they do
when $Q=0$.
They always scale as $\sqrt{\rho_b/\rho_a}$, regardless of the value
of $\w$.

In order to see how accurate our code is at the shock, we generated
shocks by evolving
general relativistic voids with $\RW=50$, $\ep_{\rm out}(\ti)/c^2=10^6$,
$\w=1/3$, $\alp=10^{-10}$, and $\rw=.25$ and $\rw=.5$.
(In addition, $C=.3$ and $\f=.01$).
After a shock formed,
we calculated $\rho_b$ by averaging $\rho$ for the first few grid points
of $\rho$ behind the shock.  In addition, we found $\Phi_b$,
$\ep_b$, $\Gamma_b$ and $n_b$ similarly.  Then, we plugged
$\rho_b$ into Eqs.({\ref{eq:Ur}})-({\ref{eq:phibr}})
to calculate the predicted values.
The percent change is plotted in
{\it Figure 13}.
These errors are expected to decrease when the artificial
viscosity is decreased, since for these runs, the number of grid points
in the shock front is $5-9$.

\newpage

\begin{table}[t]
\begin{center}
\caption{Relativistic Homogeneous convergence test for $\Delta t_{c_S} < \Delta
t_n$}
\begin{tabular}{|c|r|r|r|r|r|}
\hline
\hline
\multicolumn{1}{|c|}{ }&\multicolumn{5}{|c|}{$(L_1(q)\times
10^{6})~~/~~s(q)$}\\
\cline{2-6}
\multicolumn{1}{|c|}{$\DR$}
&\multicolumn{1}{|c|}{$n$}&\multicolumn{1}{|c|}{$R$}&\multicolumn{1}{|c|}{$U$}
&\multicolumn{1}{|c|}{$M$}&\multicolumn{1}{|c|}{$\ep$}\\
\hline
\hline
$ 20.0 $&$28.\;\;/$  --- &$2.5\;\;/$  ---&$7.6\;\;/$  ---&$18.\;\;/$
---&$2.4\;\;/$  ---\\
\hline
$ 10.0
$&$6.9\;\;/\;\;2.0$&$.62\;\;/\;\;2.0$&$1.9\;\;/\;\;2.0$&$4.4\;\;/\;\;2.0$&
$.61\;\;/\;\;2.0$\\
\hline
$ 5.00
$&$1.7\;\;/\;\;2.0$&$.15\;\;/\;\;2.0$&$.48\;\;/\;\;2.0$&$1.1\;\;/\;\;2.0$&
$.15\;\;/\;\;2.0$\\
\hline
$ 2.50
$&$.43\;\;/\;\;2.0$&$.038\;\;/\;\;2.0$&$.12\;\;/\;\;2.0$&
$.29\;\;/\;\;1.9$&$.036\;\;/\;\;2.1$\\
\hline
$ 1.25
$&$.11\;\;/\;\;2.0$&$.0095\;\;/\;\;2.0$&$.037\;\;/\;\;1.7$&
$.084\;\;/\;\;1.8$&$.0074\;\;/\;\;2.3$\\
\hline
\hline
\end{tabular}
\end{center}
\end{table}

\begin{table}[t]
\begin{center}
\caption{Relativistic Homogeneous convergence test for $\Delta t_n < \Delta
t_{c_S}$}
\begin{tabular}{|c|r|r|r|r|r|}
\hline
\hline
\multicolumn{1}{|c|}{ }&\multicolumn{5}{|c|}{$(L_1(q)\times
10^{7})~~/~~\hat{s}(q)$}\\
\cline{2-6}
\multicolumn{1}{|c|}{$\f$}
&\multicolumn{1}{|c|}{$n$}&\multicolumn{1}{|c|}{$R$}&\multicolumn{1}{|c|}{$U$}
&\multicolumn{1}{|c|}{$M$}&\multicolumn{1}{|c|}{$\ep$}\\
\hline
\hline
$ .02 $&$24.\;\;/$  --- &$2.2\;\;/$  ---&$6.7\;\;/$  ---&$16.\;\;/$
---&$2.2\;\;/$  ---\\
\hline
$ .01
$&$5.1\;\;/\;\;2.2$&$.46\;\;/\;\;2.3$&$1.4\;\;/\;\;2.3$&$3.3\;\;/\;\;2.3$&
$.45\;\;/\;\;2.3$\\
\hline
$ .005
$&$1.1\;\;/\;\;2.2$&$.10\;\;/\;\;2.2$&$.34\;\;/\;\;2.0$&$.79\;\;/\;\;2.1$&
$.094\;\;/\;\;2.3$\\
\hline
$ .0025
$&$.27\;\;/\;\;2.0$&$.024\;\;/\;\;2.1$&$.10\;\;/\;\;1.8$&$.23\;\;/\;\;1.8$&
$.016\;\;/\;\;2.6$\\
\hline
\hline
\end{tabular}
\end{center}
\end{table}

\begin{table}[t]
\begin{center}
\caption{Collapse Times for a Superhorizon-Sized relativistic void $25$ times
the Hubble radius}
\begin{tabular}{|c|l|l|l|c|}
\hline
\hline
\multicolumn{1}{|c|}{ }
&\multicolumn{3}{|c|}{$\sG/\rw\,\,\, | \,\,\, \Delta t_{\rm collapse}/\Hm$}
&\multicolumn{1}{|c|}{ }\\
\cline{2-4}
\multicolumn{1}{|c|}{$\GW$}
&\multicolumn{1}{|c|}{$\alp=10^{-7}$}&\multicolumn{1}{|c|}
{$\alp=10^{-10}$}&\multicolumn{1}{|c|}{$\alp=10^{-13}$}
&\multicolumn{1}{|c|}{$\Delta t_{\rm peak~area}$}\\
\hline
\hline
$ 6 $&$ 1.88 \,\,|\,\, .33^* $ &$ 2.05 \,\,|\,\, .047 $ &$ 2.22 \,\,|\,\, .0076
$ &$.056$\\
\hline
$ 18 $&$ 1.79 \,\,|\,\, .34^* $ &$ 1.99 \,\,|\,\, .047 $ &$ 2.17 \,\,|\,\,
.0076$ &$.041$\\
\hline
$ 50 $&$ 1.73 \,\,|\,\, .40^* $ &$ 1.94 \,\,|\,\, .049^* $ &$ 2.13 \,\,|\,\,
.0076 $ &$.020$\\
\hline
$ 100 $&---------- &$ 1.91 \,\,|\,\, .056^* $ &$ 2.11 \,\,|\,\, .0077 $&$.010$
\\
\hline
$ 200 $&---------- &$ 1.88 \,\,|\,\, .067^* $ &$ 2.08 \,\,|\,\, .0079^*
$&$.0052$ \\
\hline
\hline
\hline
\multicolumn{1}{|c|}{$\Phi_{\rm in}$ }
&\multicolumn{3}{|c|}{$\Delta t_{\rm collapse}/\Hm$}
&\multicolumn{1}{|c|}{ }\\
\hline
\hline
\multicolumn{1}{|c|}{$ \alpp^{-.25}$ }
&\multicolumn{1}{|r|}{$.44$}
&\multicolumn{1}{|r|}{$.079$}
&\multicolumn{1}{|r|}{$.014$}
&\multicolumn{1}{|c|}{ }\\
\hline
\multicolumn{1}{|c|}{$ .6\alpp^{-.3}$ }
&\multicolumn{1}{|r|}{$.33$}
&\multicolumn{1}{|r|}{$.042$}
&\multicolumn{1}{|r|}{$.0052$}
&\multicolumn{1}{|c|}{ }\\
\hline
\hline
\end{tabular}
\end{center}
\end{table}

\begin{table}[t]
\begin{center}
\caption{Collapse Times versus $\rw$}
\begin{tabular}{|c|c|c|c|c|}
\hline
\hline
\multicolumn{1}{|c|}{$\rw$}
&\multicolumn{1}{|c|}{$\RW-R_{\rm inner}$}
&\multicolumn{1}{|c|}{$\sG/\rw$}
&\multicolumn{1}{|c|}{$ \Delta t_{\rm collapse}/\Hm$}
&\multicolumn{1}{|c|}{$\pp$}\\
\hline
\hline
$ 1 $&$13$&$ 2.05 $&$ .047 $ &$ .29$\\
\hline
$ .5 $&$7$&$ 2.13 $&$ .044 $ &$ .30$\\
\hline
$ .25 $&$3$&$ 2.20 $&$ .034 $ &$ .31$\\
\hline
\hline
\end{tabular}
\end{center}
\end{table}

\begin{table}[t]
\begin{center}
\caption{Shock Location versus $\rw$ for an uncompensated SR void}
\begin{tabular}{|c|c|c|c|}
\hline
\hline
\multicolumn{1}{|c|}{$\rw$}
&\multicolumn{1}{|c|}{$R_{\rm sh}$}
&\multicolumn{1}{|c|}{$p_b$}
&\multicolumn{1}{|c|}{$\Phi_{\rm in}$}\\
\hline
\hline
$.01  $&$18.4 $&$.013$&$3.59 $\\
\hline
$.02  $&$18.4 $&$.014 $&$3.60 $\\
\hline
$.05  $&$18.5 $&$.013$&$3.57 $\\
\hline
$.1  $&$18.5 $&$.011$&$3.48$\\
\hline
$.2  $&$18.6 $&$.0085 $&$3.33$\\
\hline
\hline
Shock Tube &$18.4  $ &$.012  $ &$3.62 $\\
\hline
\end{tabular}
\end{center}
\end{table}

\end{document}